\documentclass[aps, prb,floatfix,twocolumn,superscriptadress,amsmath,amssymb,showpacs]{revtex4}
\usepackage{graphicx}
\usepackage{dcolumn}
\usepackage{bm}
\usepackage{amsmath}

\def\go{$\rightarrow$}
\def\EF{$E_{\rm F} $}
\def\kF{$k_{\rm F} $}
\def\hv{$h\nu $}

\def\G{$\Gamma$}
\def\a{$\alpha $}

\def\A-1{\AA$^{-1}$}
\def\~{$\approx$}
\def\hv{$h\nu $}
\def\CeYbCo{Ce$_{1-x}$Yb$_{x}$CoIn$_{5}$}
\def\CeCo{CeCoIn$_{5}$}
\def\YbCo{YbCoIn$_{5}$}
\def\CeRh{CeRhIn$_{5}$}
\def\TK{$T_K$}
\def\T*{$T^*$}
\def\Tc{$T_c$}

\begin{document}

\preprint{}

\title{Yb Valence Change in Ce$_{1-x}$Yb$_{x}$CoIn$_{5}$ from spectroscopy and bulk properties}

\author{L. Dudy$^{1,4}$}
\author{J. D. Denlinger$^2$}
\author{L. Shu$^{3,5}$}
\author{M. Janoschek$^{3,6}$}
\author{J. W. Allen$^1$}
\author{M.B. Maple$^3$}

\affiliation{$^1$Randall Laboratory, University of Michigan, Ann
Arbor, MI 48109-1040, USA \\ $^2$Advanced Light Source, Lawrence
Berkeley National Laboratory, Berkeley, CA, 94270,
USA\\$^3$Department of Physics, University of California, San Diego, La Jolla, CA, 92093, USA\\$^4$Physikalisches Institut, Universit\"at W\"urzburg, Am Hubland, 97074 W\"urzburg, Germany\\$^5$State Key Laboratory of Surface Physics, Department of Physics, Fudan University, Shanghai 200433, People's Republic of China\\$^6$Condensed Matter and Magnet Science group, Los Alamos National Laboratory, Los Alamos, New Mexico 87545, USA
}%

\date{\today}

\begin{abstract}

The electronic structure of \CeYbCo\ has been studied by a combination of photoemission, x-ray absorption and bulk property measurements.  Previous findings of a Ce valence near 3+ for all x and of an Yb valence near 2.3+ for $x$ $\geq$ 0.3 were confirmed.  One new result of this study is that the Yb valence for $x$ $\leq$ 0.2 increases rapidly with decreasing $x$ from 2.3 toward 3+, which correlates well with de Haas van Alphen results showing a change of Fermi surface around x=0.2.  Another new result is the direct observation by angle resolved photoemission Fermi surface maps of $\approx$ 50 $\%$ cross sectional area reductions of the \a\ and $\beta$ sheets for $x=$ 1 compared to $x=$ 0, and a smaller, essentially proportionate,  size change of the \a\ sheet for $x=$ 0.2.  These changes are found to be in good general agreement with expectations from simple electron counting.  The implications of these results for the unusual robustness of superconductivity and Kondo coherence with increasing x in this alloy system are discussed.

\end{abstract}


\maketitle

\section{\label{sect:Intro}Introduction}

\subsection{\label{subsect:GenTrans}General Overview}
Heavy fermion (HF) systems are characterized by a delicate interplay of localized and itinerant electronic degrees of freedom that is responsible for a myriad of interesting strongly correlated electron phenomena\cite{Maple10}. Here the localized $f$-electrons of Ce, Pr, Yb and various 5$f$ elements embedded in an intermetallic compound hybridize with the conduction electrons\cite{Coleman}.

The physics of the dilute limit of a single $f$-electron impurity in a metallic host is well-understood and described by the single-ion Kondo problem, where below the single ion Kondo temperature \TK\ the spins of the conduction electrons quench the local magnetic moment of the impurity via the Kondo interaction \cite{Kondo64}. HF compounds  represent the dense limit where the $f$-electron elements are arranged on a lattice, and in turn their local magnetic moments are mutually coupled through the conduction electrons by means of the Ruderman-Kittel-Kasayu-Yosida (RKKY) interaction.  The competition\cite{Doniach77}  between RKKY and Kondo interactions is often summarized in a generic phase diagram for HF materials. On one end the RKKY interaction leads to the development of long-range antiferromagnetic (AFM) order.  On the other end the Kondo interaction drives the demagnetization of the $f$-electron state, resulting in a paramagnetic HF state in which the entire lattice of $f$-electron moments collectively undergoes the Kondo effect and the $f$-electrons are delocalized into the conduction band \cite{Fisk86}.  The competing interactions can frequently be tuned by a non-thermal control parameter such as chemical composition $x$, pressure $P$ or magnetic fields $H$.  A magnetic quantum critical point (QCP) is observed when the magnetic critical temperature is suppressed to zero.

Magnetic QCPs in HF compounds have continuously attracted scientific interest because in their vicinity the Fermi-liquid paradigm is observed to break down\cite{Maple10}, frequently accompanied by the emergence of unconventional superconductivity (SC)~\cite{Pfleiderer09}. Both phenomena originate from the abundance of soft magnetic quantum fluctuations at the QCP, which in the latter case are believed to provide the pairing ``glue'' for the SC\cite{Mathur98}.

In particular, the family of tetragonal ``115'' systems has been investigated in great detail in order to disentangle the complex interplay between the heavy fermion state, unconventional SC and quantum criticality. The most prominent member of this class is \CeCo~\cite{Petrovic01b}.
\CeCo\ evinces superconductivity below a critical temperature \Tc $=$ 2{.}3 K, and is thought to be situated on the brink of an AFM QCP. This is reflected in the $T$-linear low-temperature electrical resistivity that indicates the presence of AFM quantum critical fluctuations \cite{Sidorov02}. Here magnetically mediated SC is supported by the observation of a strong spin-resonance in neutron scattering experiments \cite{Stock08}.

Apart from the SCing state, also the heavy normal state and non-Fermi liquid (NFL) properties of \CeCo\ have been studied extensively by tuning the system via rare earth substitution on the Ce site. Notably, an analysis of transport data for Ce$_{1-x}$La$_{x}$CoIn$_{5}$  introduced the notion of a coherence temperature \T*\ $\approx$ 45 K below which f-electron delocalization is supposed to proceed \cite{Nakatsuj04}. Further, for Ce$_{1-x}R_x$CoIn$_5$ it was found that both Cooper pair breaking and Kondo-lattice coherence  are uniformly influenced by magnetic and nonmagnetic rare earth ($R$) substituents. In contrast, the NFL behavior is strongly dependent on the $f$-electron configuration of the $R$ ions \cite{Paglione07}. A more recent study suggests that the introduction of small amounts of non-magnetic impurities on the Ce (Y, La, Yb, Th) and In (Hg, Sn) site generates an inhomogeneous electronic state, in which the periodicity of the Kondo lattice is disrupted by the impurities \cite{Bauer11}. This additionally results in a rapid local suppression of unconventional superconductivity.

These prior results can be contrasted with the properties of a new alloy series \CeYbCo\ for 0~$\leq$~$x$~$\leq$1 that may provide a fresh view on both the normal and the SCing state in HF compounds\cite{Capan2010, Shu2011}. Notably, it was found that the coherence temperature \T*, identified via the low-temperature electrical resistivity maximum, is essentially constant over the entire substitution range \cite{Shu2011}. This is surprising by itself, but taking into account that the single ion Kondo temperatures \TK\ for \CeCo\ and \YbCo\ differ (see, e.g., Ref.~\onlinecite{Booth2011}), it also contradicts a recent study that suggests that \T*\ and \TK\ generally scale with each other \cite{Yang2008}. The apparent stability of the electronic state is also reflected in the lattice parameters that remain nearly constant for $x$~$\leq$~0.775, after which phase separation into Yb rich and deficient phases of \CeYbCo\ occurs. Also the magnetic susceptibility is almost unaffected by the substitution of Ce with Yb. The SCing critical temperature \Tc\ decreases linearly with $x$ towards 0 K as $x$ $\rightarrow$ 1, in contrast with other HF superconductors where \Tc\ scales with \T*\ (see, e.g., Ref.~\onlinecite{Paglione07}).  Only the low-temperature NFL behavior derived from the electrical resistivity, specific heat and magnetic susceptibility varies with $x$, even though there is no readily identifiable quantum critical point.

Two different hypotheses have been put forward to explain the remarkable behavior of \CeYbCo. Based on the robustness of \Tc\ and \T*, and most of all the agreement of the observed NFL behavior with the presence of critical valence fluctuations, Shu {\it et al.} have proposed a correlated electron state having cooperative valence fluctuations of Yb and Ce \cite{Shu2011}. On the other hand, Booth {\it et al.} \cite{Booth2011} suggested from EXAFS measurements that below the Yb concentration where macroscopic phase separation takes place there is nonetheless a high degree of inhomogeneity in the form of large coexisting interlaced networks of \CeCo\ and \YbCo. It was argued that the \YbCo\ network would locally influence the physical properties of \CeCo, causing the slow suppression of \Tc. But it seems that one could equally well imagine that the consequence of such large networks could be an unchanging value of \Tc\ because the \CeCo\ and \YbCo\ networks would only influence each other at their respective surfaces. In that case, one would expect the \Tc\ in the \CeCo\ network remains constant as function of Yb concentration, while the superconducting volume fraction for the entire sample would decrease. We also note that for the samples studied by Booth {\it et al.} a change in the distances of nearest neighbor ions has been observed for $x$ \~ 0.4 using EXAFS. This may be interpreted as a local precursor of the macroscopic phase separation that has been observed by Shu {\it et al.} at $x$ \~ 0.8.

The implications of recent thin film studies are unclear for the issue of homogeneity.  It was found for epitaxial superlattices of \CeCo/\YbCo\ \cite{Mizukami2011} that \Tc\ is suppressed by $x$ \~  0.2, essentially like the behavior for other ($R$) substituents.  For thin films of \CeYbCo\ \cite{Shimozawa2012} \Tc\ is suppressed by $x$ \~  0.4, which is, on the one hand, smaller than for bulk crystals, but on the other hand, still larger than for other ($R$) substituents.  One possible interpretation could be that the films are more homogeneous than the bulk crystals and thus show \Tc\ suppression more quickly.  But the more likely possibility is that the thin films constitute an essentially different materials system from the bulk crystals, e.g., owing to the effect of the interaction of the film with the substrate.  For the thin films of \CeYbCo\ it was found \cite{Shimozawa2012} that the in-plane lattice parameter is expanded slightly because it is in registry with the substrate and does not change with $x$, whereas the out-of-plane lattice parameter varies linearly with $x$ between the values for \CeCo\ and \YbCo, quite different from the behavior of the bulk crystals. The comparison of thin films to bulk crystals is further complicated by the recent finding that thin films are extremely sensitive to air and degrade quickly upon exposure \cite{Scheffler2013}.

Both hypotheses, the cooperative valence fluctuations as well as the coexisting interlaced networks, have interesting implications and justify a more detailed microscopic investigation of the electronic structure, and in particular the Ce and Yb valences. The possibility of critical valence fluctuations within an extended SCing phase is remarkable in the context of recent studies in which CeRhIn$_{5}$~\cite{Watanabe10} and CeCu$_{2}$Si$_{2}$~\cite{Yuan03} were suggested as candidates for valence-fluctuation mediated SC. Further a recent study of the transport properties of CeRhIn$_{5}$ under hydrostatic pressure found that scattering of the charge carriers near the AFM QCP is isotropic, in contrast to expectations for a classical AFM QCP.  This finding was interpreted as a signature of coexisting critical degrees of freedom in both spin and charge channels \cite{Park08} that could be a source of SC pairing.  On the other hand SC in interlaced networks of \CeCo\ and \YbCo\ is of interest in view of a current proposal that unconventional SCs in the vicinity of AFM may be generally electronically textured \cite{Park2012}. Finally, the $x$-dependence of \T* has been addressed in a recent theoretical work which demonstrates that the onset of coherence is strongly affected by the degree of correlations between impurity sites \cite{Dzero2012}.

\subsection{Issues of valence and electron counting}\label{subsec:ecounting}

With the proposal of cooperative valence fluctuations, the
$x$-dependencies of the Ce and Yb valences are of immediate interest.
From spectroscopic information and analysis of bulk properties the
$x$~$=$~0 compound is known to have essentially trivalent Ce (4$f^{1}$)
and the uncorrelated behavior of the $x$~$=$~1 compound might suggest
divalent Yb (4$f^{14}$).  But the picture of cooperative valence
fluctuations in the alloy would require intermediate valence Yb.
This picture has been supported by a report from Booth $\it{et~al.}$
\cite{Booth2011} from X-ray absorption spectroscopy (XAS) at the Yb
and Ce L$_{III}$ edges that the Yb and Ce valences are \~ 2.3 and \~
3.1, respectively. Ref. \onlinecite{Booth2011} also finds these
valences to be essentially independent of $x$. The $x$~$=$~1 compound can
then be interpreted\cite{Booth2011} as intermediate valence with the 4$f^{13}$
magnetic moment quenched on such a high energy scale \TK\ that Curie
Weiss behavior cannot be seen.  \TK\ was estimated in Ref. \onlinecite{Booth2011} to be larger than 6000 K.

There are also important questions of electron counting and the
implications for the volume contained by the Fermi surface (FS).  It
must be kept in mind that if a local moment is quenched, e.g., by the
Kondo effect, then the electrons producing the moment must be
included in the Fermi surface (FS).  Thus for Ce$^{3+}$ with its
magnetic moment Kondo quenched, the FS volume is based on an atomic
configuration [Xe]4$f^{1}$5$d^{1}$6$s^{2}$ having 4 delocalized electrons/Ce,
the same as if Ce were formally Ce$^{4+}$
[Xe]4$f^{0}$5$d^{2}$6$s^{2}$. For \CeCo\, de Haas van Alphen (dHvA)
measurements \cite{Settai2001, Hall2001, Shishido2002}
at low temperatures have found that the 4$f$ electron is
included in the Fermi surface although angle resolved PES (ARPES)
performed at \~ 20K has found the FS expected in LDA band
calculations with the Ce 4f electron confined to the core, i.e.
[Xe][4$f^{1}$]5$d^{1}$6$s^{2}$ with 3 electrons/Ce  going into the
FS.  \cite{JD,Koitzsch2009}  Analogously for Yb$^{3+}$ with 4$f^{13}$, if the magnetic moment of
the 4$f$ hole is quenched, the Fermi surface must contain the 4$f$ hole
and so its volume will be the same as though Yb were formally
divalent [Xe][4$f^{14}$]6$s^{2}$, i.e., the FS volume would contain 2
electrons/Yb.  Because of its large \TK\ this situation is expected up to very high
temperatures for \YbCo .  Thus, no matter whether or not the 4$f$
electron in \CeCo\ is localized at the measurement temperature, one
expects that in comparing the $x$~$=$~1 and the $x$~$=$~0 compounds,
characteristic hole FS features will tend to expand and
characteristic electron FS features will tend to shrink with increasing x.

Recent dHvA experiments \cite{Polyakov2012} found that the FS of \YbCo\ is indeed much different from that of \CeCo.  The measured frequencies are in reasonable agreement with an LDA calculation for \YbCo, which gives an Yb valence of 2.3, the same as found in XAS \cite{Booth2011}, and shows reduced volume relative to that of $x$~$=$~0.  For example the frequencies assigned to two prominent electron FS features centered on the M-A line in $k$-space and known as $\alpha$ and $\beta$ from LDA \cite{Settai2001,Elgazzar04} and dHvA \cite{Settai2001, Hall2001, Shishido2002} studies for $x$~$=$~0, decrease markedly from $x$~$=$~0 to $x$~$=$~1.  It is also of note that the dHvA effective masses for \YbCo\ are relatively small, in the range 1.0 $m_e$ to 1.5 $m_e$. These small masses are consistent with the very large value of \TK\ implied \cite{Booth2011} by the Yb valence of 2.3.   For $x$~$=$~0.1, the dHvA frequencies and masses are unchanged from those of $x$~$=$~0 and for $x$~$=$~0.2 there appear frequencies characteristic of both $x$~$=$~0 and $x$~$=$~1.  For $x$~$=$~0.55, the next highest value for which dHvA data were obtained, and for higher values $x$~$=$~0.85 and 0.95, the frequencies and masses that could be observed are generally like those that are found for $x$~$=$~1, with the $\alpha$ frequencies essentially unchanged and the $\beta$ frequencies changing slightly.  Thus dHvA shows a rather abrupt change of electronic structure around $x$~$=$~0.2.  In contrast, Ref. \onlinecite{Booth2011} reported from ARPES that the electronic structure along the \G -M line is essentially invariant with $x$, including $x$~$=$~1.

\subsection{Present work and Organization of the Paper}\label{subsec:resultsorg}

In the present work we determine the Ce valence from XAS at the Ce
M$_{4,5}$ edges and the Yb valence from 4$f$ electron x-ray
photoemission spectroscopy (XPS).  In agreement with Ref. \onlinecite{Booth2011},
 we find that the Ce valence is near 3+ and
essentially independent of $x$. The Yb valence for $x$~$=$~1 and decreasing
to \~ 0.3 is \~ 2.3, also in agreement with the finding of Ref.
\onlinecite{Booth2011}. But as $x$ decreases further and approaches 0
the valence increases to nearly 3+.  We also report and analyze the
$x$-dependence of the alloy magnetic susceptibility for temperatures
where it exhibits Curie-Weiss behavior, and we introduce a simple
model to inter-relate the effective moment and the Ce and Yb
valences.  Under the assumptions that Kondo effects can be ignored
and that the Ce valence is essentially 3+, we infer Yb valences in
very good agreement with the values found from XPS, including their
tendency to increase toward 3+ for small $x$.  Thus the rather abrupt change of electronic structure found in dHvA can now be seen as resulting from a change of Yb valence.

We present the $k$-dependent electronic structure and FS of \CeCo\ and \YbCo\ throughout the Brillouin zone as measured using variable photon energy ARPES.  In contrast to the results of Ref. \onlinecite{Booth2011} we find along the \G -M line a large difference of electronic structure for $x$~$=$~0 and $x$~$=$~1. In particular the sizes of the $\alpha$ and $\beta$ sheets decrease markedly from $x$~$=$~0 to $x$~$=$~1, in good qualitative agreement with the general expectations from electron counting set forth above and with the dHvA results \cite{Polyakov2012}.   But in disagreement with dHvA is the observation of a smaller, essentially proportionate,  size change of the \a\ sheet for $x=$ 0.2. The somewhat columnar shapes of these FS features have drawn attention in connection with the idea that the layered crystal structure may be important for the SC of the $x$~$=$~0 compound.

The remainder of the paper is organized as follows.  The experimental details on the sample preparation, bulk property measurements and spectroscopic measurements are summarized in section ~\ref{sec:expdetails}. Section ~\ref{sec:spectroscopy} presents the various spectroscopic studies and the data analysis used to determine the Yb and Ce valences. In section ~\ref{sec:bulk} we relate the Ce and Yb valences to the dependence of the magnetic susceptibility and the unit cell volume on the Yb concentration $x$. Our ARPES results for $x$~$=$~0, 0.2 and 1 are presented in section ~\ref{sec:arpes}. We end with a summary and our conclusions in section ~\ref{sec:conclusion}.

\section{Experimental Details}\label{sec:expdetails}
Single crystals of \CeYbCo\ were grown using  an indium self flux
method \cite{Petrovic01a,Zapf01}. High purity elements (Ce, 3N; Yb,
3N; Co, 3N; In, 4N) were placed in alumina crucibles and heated in
quartz tubes with 150 psi argon gas. The heating schedule consisted
of an initial ramp at 50~$^\circ$C/hr to 1050~$^\circ$C, a dwell at
1050~$^\circ$C for 72 hours, and a two-stage cooling process to
avoid forming crystals of CeIn$_3$ $-$ first a rapid cooling from
1050~$^\circ$C to 800~$^\circ$C followed by a slow cool to
450~$^\circ$C, where the excess flux was spun off in a
centrifuge.\cite{Petrovic01a} The resulting crystals were
characterized with x-ray powder diffraction (XRD) and energy
dispersive x-ray (EDX) analysis in order to verify both the correct
structure and composition. The magnetization $M_{ab}$ of the
crystals was measured as a function of temperature for 2~K
$\leqslant T \leqslant$ 300~K using a Quantum Design superconducting
quantum interference device (SQUID) magnetometer in a magnetic field
$H$~$=$~5000 Oe applied parallel  and perpendicular to the basal
tetragonal plane.

The spectroscopic measurements XAS, XPS and ARPES were performed at
undulator beamline 7.0.1 of the Advanced Light Source (ALS)
synchrotron. XAS was measured using total electron yield (TEY), by
simultaneously measuring the sample current and the reference of a
Ni mesh located before the sample in the monochromator.  A Scienta
R4000 electron spectrometer with 2D parallel detection of electron
kinetic energy and angle in combination with a highly-automated
six-axis helium cryostat goniometer was used to acquire Fermi
surface (FS) and electronic structure maps with a wide $>$30
$^{\circ}$ angular window covering multiple Brillouin zones (BZs).
The measurements were performed with pressure between 8x10$^{-11}$
mbar and 1x10$^{-10}$ mbar. The samples were cleaved in situ by
pushing against a post which was glued on the sample surface by
epoxy adhesive. The cleavage temperature was between 20 K and 25 K,
essentially equal to the measurement temperature, which was 25 K for
\CeCo~ and 20K for \YbCo~. The position of the Fermi energy and the
energy resolution were determined by measuring a gold foil adjacent
to and in good thermoelectrical contact with the sample.  Before
measuring, the gold foil was scraped  $\it{in\-situ}$ to obtain a clean
surface.  Within about 100 $\mu$m all spectroscopic data were
collected on the same spot of the cleavage plane, which showed no
visible inhomogeneities either $\it{in\-situ}$ or afterwards in images in
both an optical microscope and an electron microscope.

XAS measurements had a resolution of 250 meV. For XPS, the overall
energy resolution was set to 140 meV FWHM for photon energies of \hv
$=$ 550\,eV and 250\,meV for photon energies of \hv $=$ 865\,eV. For ARPES
the total energy resolution of the analyzer and exciting photons
varied from ~30\,meV at \hv $=$ 80\,eV to ~45\,meV at \hv $=$ 200\,eV, and
the angular resolution of ~0.3\,$^{\circ}$ corresponds to a parallel
angular momentum resolution range of 0.024 ${\rm \AA}^{-1}$ to 0.037
${\rm \AA}^{-1}$.  Detector angular distortions are corrected using
calibration data acquired with a slit array placed between the
sample and analyzer lens. Angular and photon-dependent Fermi-energy
maps were extracted with an energy width of 50\,meV. The value of k
perpendicular to the sample surface (k$_z$) could be selected by
varying the photon energy, as verified and calibrated from repeating
features in $k_z$-$k_x$ maps using a standard method
\cite{Himpsel1983} that approximates the photoelectron dispersion by
a free electron parabola and an "inner potential" $V{_0}$ to
characterize the surface potential discontinuity.  A $V{_0}$  value
of 11.9 $\pm$ 0.6\,eV best describes the repeating features in the
data.

\begin{figure}[h]
  \begin{center}
  \includegraphics[width=0.5\textwidth]{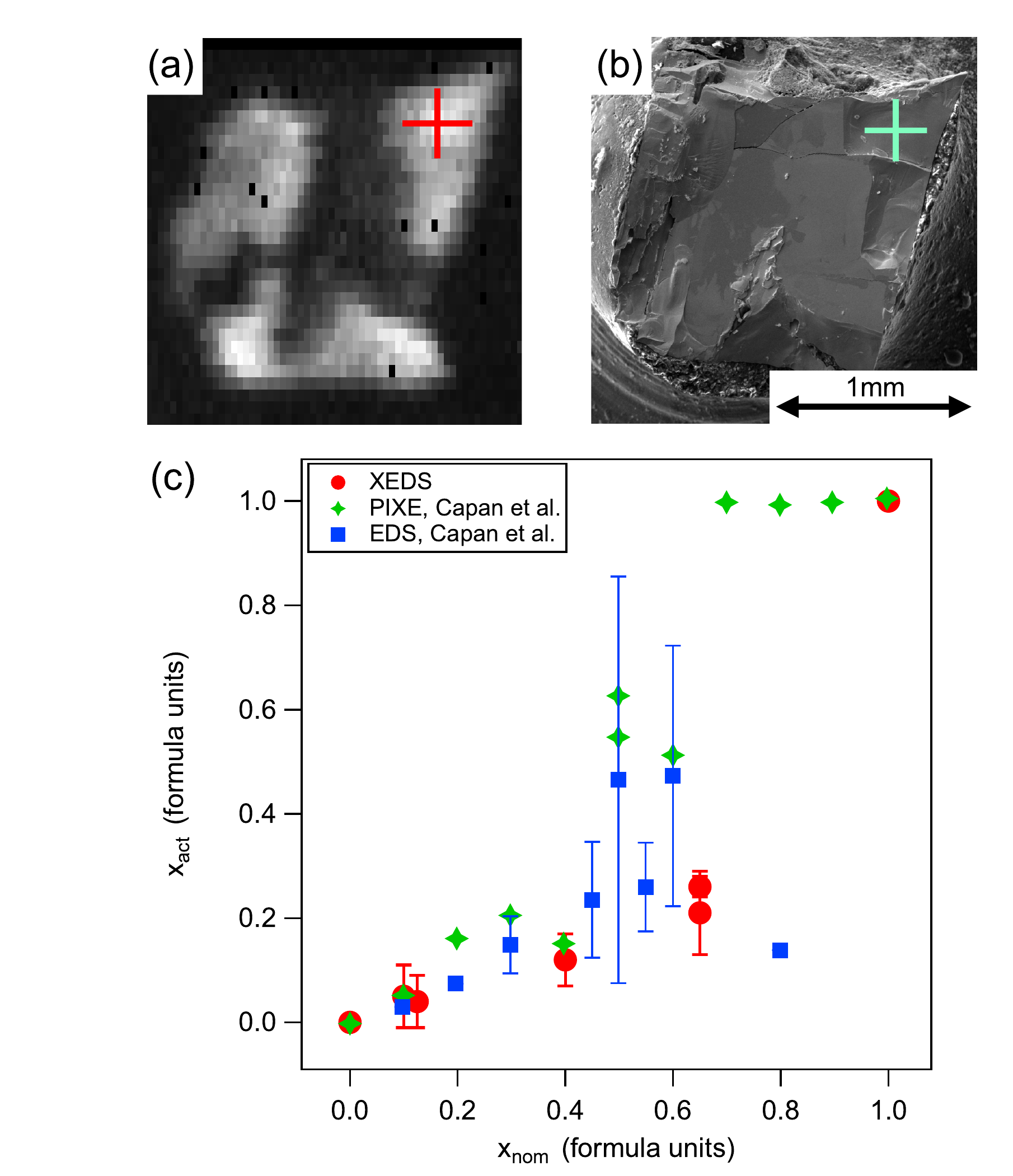}
  \caption{(a) Photoelectron intensity in a window of 200 meV around the Fermi-energy in dependence of the x- and y-sample position perpendicular to the sample-analyzer axis. Brighter color means higher intensity of the photoelectrons. The cross marks the position where the photoelectron spectroscopy was performed. (b) Image of the same sample in the scanning electron microscope. The cross marks the position where the XEDS analysis was performed. (c) Nominal Yb-concentration vs. the actual Yb-concentration as obtained by the XEDS analysis (please compare with Tab. \ref{tab:EDS}). The results are compared with the results of Capan et al. \cite{Capan2010}. }
  \label{fig:EDS}
  \end{center}
\end{figure}

The actual compositions $x_{act}$ of the \CeYbCo\ samples were determined by X-ray energy dispersive spectrometry (XEDS) analysis.  Although $x_{act}$ can vary considerably from the nominal composition $x_{nom}$, as described further below,  nonetheless it is possible to find samples in which the two are similar and such samples were carefully selected for obtaining the susceptibility data reported here and in Ref. \onlinecite{Shu2011}.  The XEDS analysis for these samples was performed at the University of California at San Diego.  The XEDS analysis of the samples used for the synchrotron electron spectroscopy was performed at the Electron Microbeam Analysis Laboratory at the University of Michigan.  Since the synchrotron spectroscopy is performed with a small photon spot on a selected region of a cleaved surface the XEDS analysis was performed after the synchrotron spectroscopy was done and great care was taken to obtain the composition at the same location on the sample as that where the synchrotron spectroscopy was performed.  Using the focused probe of the scanning electron microscope it is possible to determine the composition of the sample with a resolution of a few $\mu$m (the actual value depends on the average atomic number of the sample and the accelerating voltage of the microscope (30kV in this case)).  Therefore by this method we can detect inhomogeneities on the micro-scale but not the nano-scale range. As the photon beam-spot for the synchrotron spectroscopy was much larger, roughly 50-100 $\mu$m, we checked the micro-scale homogeneity by doing XEDS analysis around the measured position, at two or three points roughly within a circle of 100-200 microns diameter.  We detected no significant changes in the Yb compositions, determined as we describe in detail next.

Fig. \ref{fig:EDS} (a) and (b) illustrate the procedure for a sample with $x_{nom}=0.4$. In the synchrotron experiment we obtain the real space sample surface ``XY'' map displayed in (a) by measuring the photoelectron intensity in a window of 200 meV around the Fermi-energy while scanning the sample position in the X- and Y-directions perpendicular to the sample-analyzer axis. Brighter color means higher photoelectron intensity. We then compare the XY-map with the scanning electron microscope image of the sample and identify the position where we performed the electron spectroscopy. This enables us to perform the XEDS analysis at very much the same position. After obtaining the X-ray fluorescence spectrum, the composition analysis was performed by using the intensity of the In L-lines, the Ce L-lines, the Co K-lines and the Yb-L lines. The first step is to perform a so-called ``standard-free'' analysis that determines the intensities of the multiple lines of the spectrum and then applies a correction in order to account for the deviations of the spectrum of a pure element relative to that of the element residing in the matrix of other elements such as in \CeYbCo.  We used the so-called ZAF method \cite{ZAF}.  This method includes an atomic number correction (Z) which estimates the backscattering and stopping power of the incident electron beam, an absorption correction (A) which corrects for absorption within the matrix, and a correction for fluorescence (F) within the matrix. The corrected spectroscopic intensities are then used to obtain relative elemental compositions subject to the constraint that there are 7 atoms per formula unit.  For In and Co compositions so obtained were typically near to the expected values of 5 and 1, respectively, but with some outlier values that for In were at most 2.6\% different from 5.  The magnitude of the maximum outlier discrepancy for the Co concentration was somewhat less than that for the In composition, resulting in a maximum percentage deviation of 9\%.  We take these discrepancies as an indication of the uncertainties inherent in the technique. The standard free Ce and Yb compositions so obtained for the Ce ($y_{sf}$) and Yb ($x_{sf}$) concentrations of samples with various $x_{nom}$ values are listed in Table \ref{tab:EDS}. In the next step the end-members \CeCo\ and \YbCo\ were used as standards in order to account for possible systematic errors in the ZAF procedure,  i.e. we would expect our true $y$=1 in \CeCo\ and our true $x$=1 in \YbCo. While the Ce component requires no correction (as \CeCo\ gives $y_{sf}$=1), the discrepancy for \YbCo\ results in the correction formula $x_{act}=x_{sf}/0.92$. For simplicity in the rest of the paper we use $x$ to mean the actual concentration $x_{act}$ so determined.  We show error bars for $x$ estimated as $\Delta x_{act}=\left|x_{act}+y_{sf}-1\right|$, reflecting the assumption that all Yb-atoms substitute for Ce-atoms and, therefore, the determination of ($1-y_{sf}$) should be also an equally valid determination of the Yb content.  In  Fig. \ref{fig:EDS} (c), we plot the nominal Yb-concentration vs. the actual Yb-concentration. One can see that, beside $x_{nom}=1$, all samples in our spectroscopic study are below the phase separation region which is between $x_{nom}=0.775$ and $x_{nom}=1$. We compared our results (circles) with the results of Capan et al. \cite{Capan2010}, who performed XEDS on single crystals and also proton-induced X-ray emission microprobe (PIXE) on a mosaic of crystals from the same batches. One sees clearly a good agreement between our results and the ones of Ref. \onlinecite{Capan2010} except for two outlier $x_{nom}$ values of the latter data.  This agreement with independent results from two other techniques encourages us to have confidence in our procedure.  We note that the general conclusions drawn in our paper are made with full cognizance of our error bars and do not depend on whether or not the systematic Yb correction described above was made.

\begin{table}[htb]
\begin{center}
\begin{tabular}{|c||c|c|c|c|c|c|c|c|}
\hline
Element & \multicolumn{7}{c|}{Nominal composition $x_{nom}$} \\
  & 0 & 0.1 & 0.125 & 0.4 & 0.65 & 0.65 & 1 \\
  \hline \hline
Ce content $y_{sf}$ & 1.00 & 1.01 & 1.01 & 0.93 & 0.77 & 0.87& 0.00\\ \hline
Yb content $x_{sf}$ & 0.00 & 0.04 & 0.04 & 0.11 & 0.24 & 0.19& 0.92\\ \hline
\hline
$x_{act}$& 0.00 & 0.04 & 0.04 & 0.12 & 0.26 & 0.21 & 1.00\\
\hline
$\Delta x_{act}$& 0.00 & 0.05 & 0.05 & 0.05& 0.03& 0.08& 0.00\\
\hline
\end{tabular}
\end{center}
\caption{Results of the XEDS analysis.}\label{tab:EDS}
\end{table}

\section{Ce and Yb valences from spectroscopy}\label{sec:spectroscopy}

\subsection{XAS for Ce valence}

\begin{figure}[h]
  \begin{center}
  \includegraphics[width=0.43\textwidth]{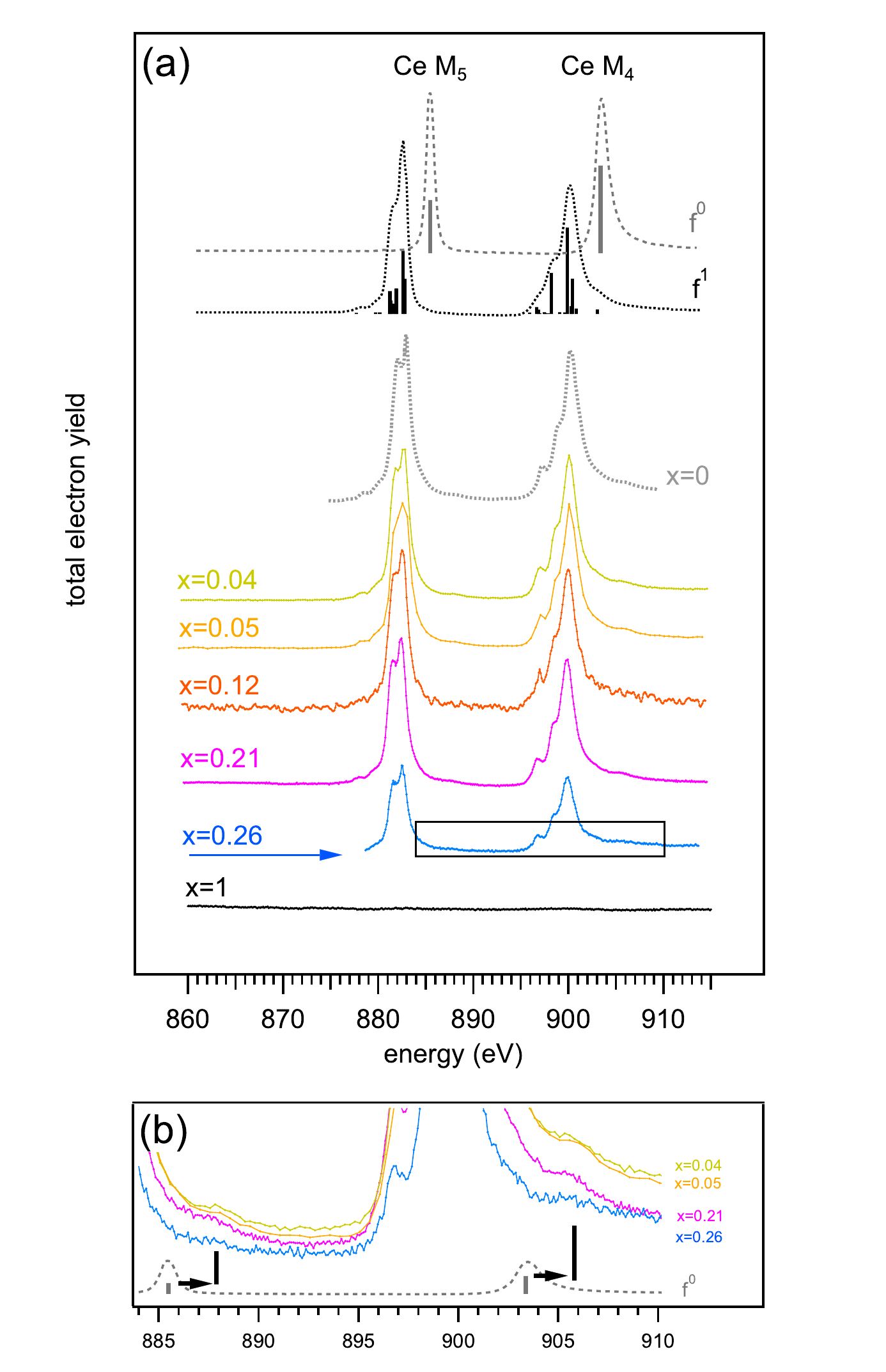}
  \caption{(a) X-ray absorption measurements at $T$~$=$~20 K near the Ce M$_{4}$ and M$_{5}$ edges show that Ce is essentially trivalent.  The topmost curves, labeled on the right side with  $f^{0}$ or $f^{1}$ are multiplet calculations \cite{Thole1985}.
  Below are measured spectra for the series \CeYbCo.  The curve for $x$~$=$~0 is from Ref. \onlinecite{Willers2010}. The box in (a) indicates the region displayed in (b) which is the magnification of the low intensity region where the $f^{0}$-component shows up. }
  \label{fig:XAS}
  \end{center}
\end{figure}

In order to determine the change of the Ce valence upon doping, we
performed TEY XAS near the Ce M$_{4}$ and M$_{5}$ edges. In this
experiment we look at the 3d$^{10}$4f$^{0} $\go\ 3d$^{9}$4f$^{1}$ and
3d$^{10}$4f$^{1}$ \go\ 3d$^{9}$4f$^{2}$ absorption lines, which are
further separated accordingly to the core hole being 3d$_{5/2}$
(M$_{5}$) or 3d$_{3/2}$ (M$_{4}$).  Because the d core-hole
interacts strongly with the promoted f-electron, this absorption
results in strong excitonic lines at energies below the true 3d \go\
6p absorption edges, with characteristic structure due to multiplet
splittings of the final states.  In the two topmost curves of Fig.
\ref{fig:XAS} (a), we show an atomic multiplet calculation
\cite{Thole1985} for the initial state (final state) being $f^{0}$
($f^{1}$) or $f^{1}$ ($f^{2}$).

A mixed valence system shows absorption lines of both valences.  The
intensity ratio R $=$ I($f^{0}$)/[I($f^{0}$)+ I($f^{1}$)], where
I($f^{0}$) and I($f^{1}$)] are the integrated intensities for lines
with initial states $f^{0}$ and $f^{1}$, respectively, is a measure
of the initial state $f^{0}$ component (1-$n_{f}^{\rm Ce}$), where
$n_{f}^{\rm Ce}$ is the Ce 4$f$ occupancy. There are, however, the
following caveats.  First, within the framework of the Anderson
impurity model, R underestimates (1-$n_{f}^{\rm Ce}$) due to the
mixing of the $f^{1}$ and $f^{2}$ final states through the
hybridization of the $f$-states with the conduction electron states
\cite{Gunnarsson1983,Fuggle1983}.  Second, TEY detection has a
contribution from the surface, which can have smaller (1-$n_{f}^{\rm
Ce}$) than the bulk because the hybridization can be reduced and the
4$f$ binding energy increased relative to the bulk. Generally, the
total fluorescence yield is more bulk sensitive than TEY and also
tends to enhance the $f^{0}$ peak due to the bulk self-absorption of
the much stronger Ce $f^{1}$-component. The greatest sensitivity to
the $f^{0}$ component is achieved through XAS on L$_{III}$ edges and
resonant x-ray emission \cite{Dallera2004}. Thus really precise
absolute values for the Ce valence are not accessible here.
Nonetheless the M$_{4,5}$ spectra are known to be a very reliable
qualitative guide to (1-$n_{f}^{\rm Ce}$) and are very accurate for the main purpose here of detecting a relative change with $x$ if it exists.

The lower curves in Fig. \ref{fig:XAS} (a)  show experimental XAS
results, all at $T$~$=$~20 K, for the series \CeYbCo. The curve for $x$~$=$~0 is
from Ref. \onlinecite{Willers2010} and indicates therefore
reproducibility of the results. Our spectra were normalized at the
background from \hv $=$ 910 \,eV to \hv $=$ 915 \,eV. The spectrum for $x$~$=$~1
shows expectedly no Ce signal. For $x<$ 1, the spectra are
dominated by the $f^{1}$-component with only a small
$f^{0}$-satellite. In Fig. \ref{fig:XAS} (b), we show a
magnification of the low intensity region. At about \hv $=$ 888 \,eV and
\hv $=$ 906 \,eV, there are little humps due to the $f^{0}$-component.
Also one can see in that magnification that the (dotted) curve from
the atomic multiplet calculation for $f^{0}$ has to be shifted by
about +2.5 \,eV to account for increased screening in the solid state. Overall, the
measured curves do not strongly change with $x$.  We can safely
conclude, in agreement with Ref. \onlinecite{Booth2011}, that Ce is
essentially trivalent and unchanging for the whole measured series. As a
quantitative measure of a possible valence change of the Ce, we
determined the intensity of the $f^{1}$-component by a fit with five
Gaussians for M$_{5}$ and four Gaussians for M$_{4}$. Similarly, we
determined the $f^{0}$-component by one Gaussian each for M$_{5}$
and M$_{4}$. Thereby we find at $T$~$=$~20 K that the value of R is
essentially constant at 0.04 $\pm$ 0.04 and 0.1 $\pm$ 0.12  for
M$_{5}$ and M$_{4}$, respectively.

\begin{figure*}
  \begin{center}
  \includegraphics[width=0.85\textwidth]{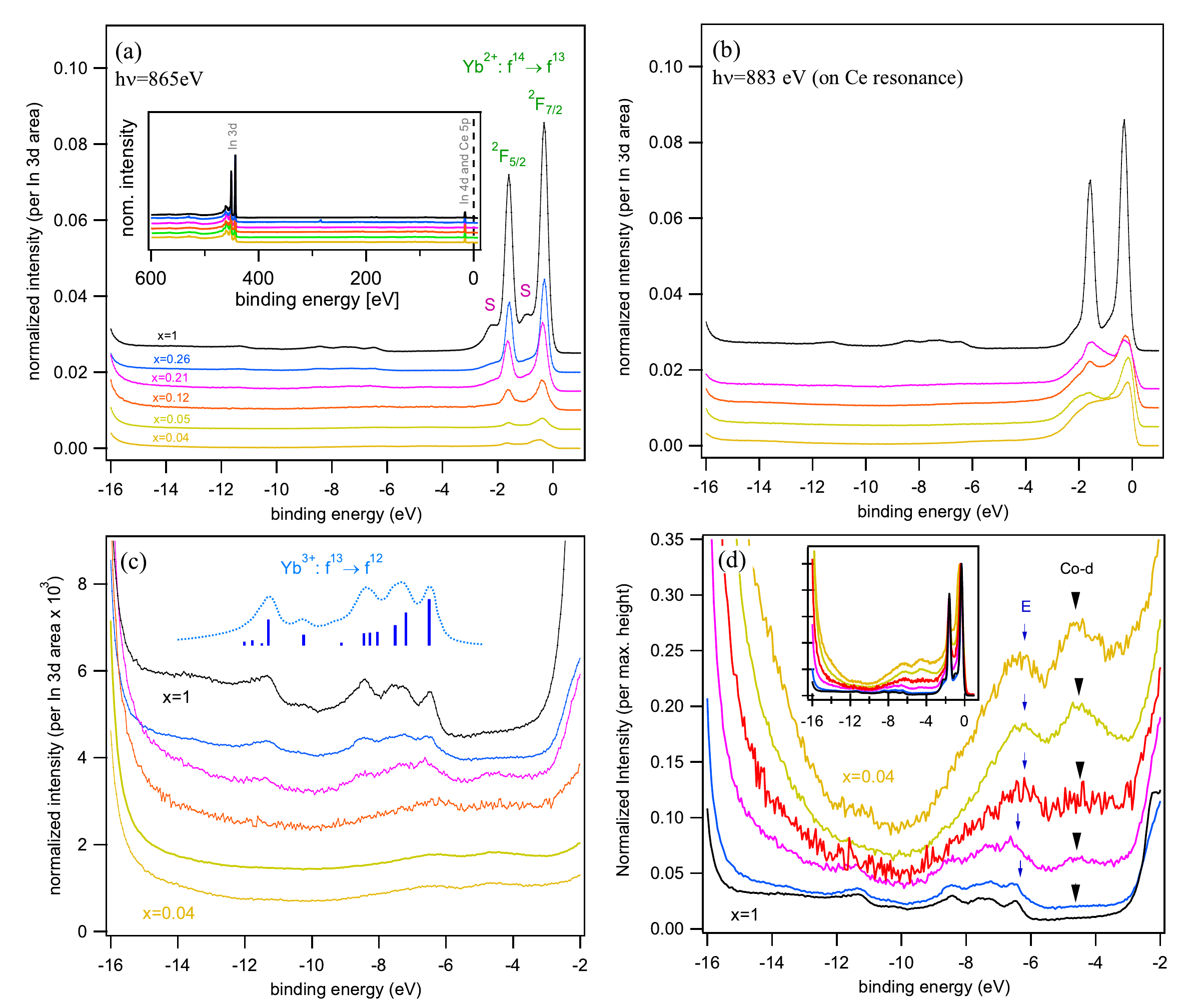}
  \caption{XPS spectra of the \CeYbCo~ samples at T=20 K. The spectra are sorted by their nominal Yb content.
  Except for the reversed order in (d), the topmost spectrum is for x=1 and the bottom one is for x=0.1. (a) Spectra of the valence band region near the Fermi energy taken at \hv =865\,eV . The doublet labeled $^2$F$_{5/2}$ - $^2$F$_{7/2}$  is due to
  bulk Yb$^{2+}$. The two peaks due to surface Yb$^{2+}$ are labeled 'S'. The inset of (a) shows a wide scan over a large
  binding energy region. This spectrum is dominated by the In 3d doublet which was used to normalize the spectra.
  Near the Fermi-energy, a peak originating from In 4d and Ce 5p is visible.
  (b) Spectra of the same valence band region as in (a) but on the Ce resonance at photon energies of \hv =833 \,eV. This
  increases the visibility of the Ce component in the spectra. Panel (c) shows a magnification of the
  region in the spectra of (a) where the Yb$^{3+}$ component can be detected. The topmost curve is a theoretical
  calculation for the Yb$^{3+}$ 5f$^{13}$ \go 5f$^{12}$  spectrum from Ref. \onlinecite{Gerken1983}. (d) is the same
  spectrum as (c) but normalized to the maximum of the $^2$F$_{7/2}$-peak as shown in the inset.
  The peak maximum originating from Co-d weight and the low energy edge "E" of the
  Yb$^{3+}$ 5f$^{13}$ \go 5f$^{12}$  multiplet are marked.}
  \label{fig:XPS}
  \end{center}
\end{figure*}

\subsection{4f PES for Yb valence}

In order to elucidate the valence of the Yb, we analyze 4$f$ XPS
spectra. At the beginning of our measurements, we routinely confirm
the absence of oxygenated surfaces by taking scans over a large
binding energy region that includes the core-states as shown in the
inset of Fig. \ref{fig:XPS} (a). These wide scans also offer the
opportunity to normalize our valence-band spectra by assuming that
each photon energy the area of the In-3$d$ peaks is constant for the
whole series of \CeYbCo.  We show such normalized valence band
spectra for the photon energy of \hv =865 \,eV in Fig. \ref{fig:XPS}
(a). The data are stacked with a constant shift in intensity and are
ordered with increasing Yb concentration from bottom to top, i.e.
the lowest is x=0.04 followed by x=0.05, 0.12, 0.21, 0.26, 1. The first
notable feature, which we label in the spectra, is the final state
$^2$F$_{5/2}$ - $^2$F$_{7/2}$ doublet of the Yb$^{2+}$ photoemission process
4$f^{14}$ \go 4$f^{13}$.  This feature is quite intense because the
large number of fourteen Yb $f$-electrons causes a much stronger
photoemission signal compared to those of Ce, In, or Co. Tuning the
photon energy to the M$_{5}$ resonance at \hv =883 \,eV (compare also
with Fig. \ref{fig:XAS}) allows us to see the Ce-weight as shown in
Fig. \ref{fig:XPS} (b).  As expected we see that this Ce weight
decreases upon adding more Yb although the Yb$^{2+}$ doublet is
still very strong as seen by comparing the intensity for the x=1
sample, which certainly has no Ce weight, with the Ce-signal
enhanced intensity for all spectra for x $<$ 1. Going back to Fig.
\ref{fig:XPS}(a), the second notable feature is two lower intensity
peaks which are marked by 'S' above them. These two peaks show the
same energy separation between each other as the separation of the
$^2$F$_{5/2}$ - $^2$F$_{7/2}$ doublet described above. Thus they are also
due to Yb$^{2+}$ but instead of coming from the bulk this signal
comes from the surface. The absence of ligand-atoms causes a
stronger screening of the $f$-levels and therefore the spectrum shifts
down in energy.  Such surface shifted Yb$^{2+}$ peaks are well
known.

The third feature in the spectrum, finally, is of lower intensity
and can be better seen in Fig. \ref{fig:XPS} (c), which shows a
magnification of the interesting region.  Although the structure
consists of many peaks their mutual origin is from Yb$^{3+}$ for
which the photoemission process 5$f^{13}$\go 5$f^{12}$ gives a final
state multiplet of thirteen lines. The topmost curve serves as a
fingerprint for Yb$^{3+}$ as it represents a calculation of this
multiplet structure using intermediate coupling \cite{Gerken1983}.
Counting the lines of this multiplet, one finds only twelve. The
missing line is from a $^{1}$S state at \~16.3 \,eV binding energy,
which causes it to be merged with the strong signal of the In 4d and
Ce 5p doublets. For our later quantitative discussion, it is good to
note that, according to the calculation, this line has only about
0.09/13 \~ 0.7 \% of the total intensity of the multiplet and is
therefore quite negligible.  We notice in Fig. \ref{fig:XPS} (c)
that, as we would expect,
 the intensity of the Yb$^{3+}$-signal decreases as the concentration of Yb
 decreases. This however does not necessarily mean that the Yb valence
has to change.

Before we start to evaluate the Yb valence qualitatively, we note
that there is a relatively easy method to graphically visualize the
Yb valence by normalizing the data in a different way. In Fig.
\ref{fig:XPS} (d) we show the same spectra as in (c) and (a), at \hv
=865 \,eV, but with another normalization. This normalization just
divides the spectrum by the maximum intensity of the $^2$F$_{7/2}$ peak.
For x $<$ 1 there is Ce-weight buried under these peaks. Thus we are
overestimating the Yb$^{2+}$ component for smaller x. There is one
feature which is more revealed as the absolute intensity of the
Yb-related features is more and more decreased. It is marked with an
arrow and it stems from the Co d-states. This growing of the Co-d
related weight shifts the edge "E" of the foremost part of the
Yb$^{3+}$ multiplet up. Taking the height between the edge "E" and
the dip located between E and the Co-d peak, we can qualitatively
state that for x=1 the Yb$^{2+}$ component is strong. We can also
clearly see that the Yb$^{3+}$ component has a lower intensity for
x=1 and x=0.26 than it has for x $<$ 0.26. This is even true without
taking account of the the fact that the normalization overestimates
the Yb$^{2+}$ component for low x.

\begin{figure}[h]
  \begin{center}
  \includegraphics[width=0.45\textwidth]{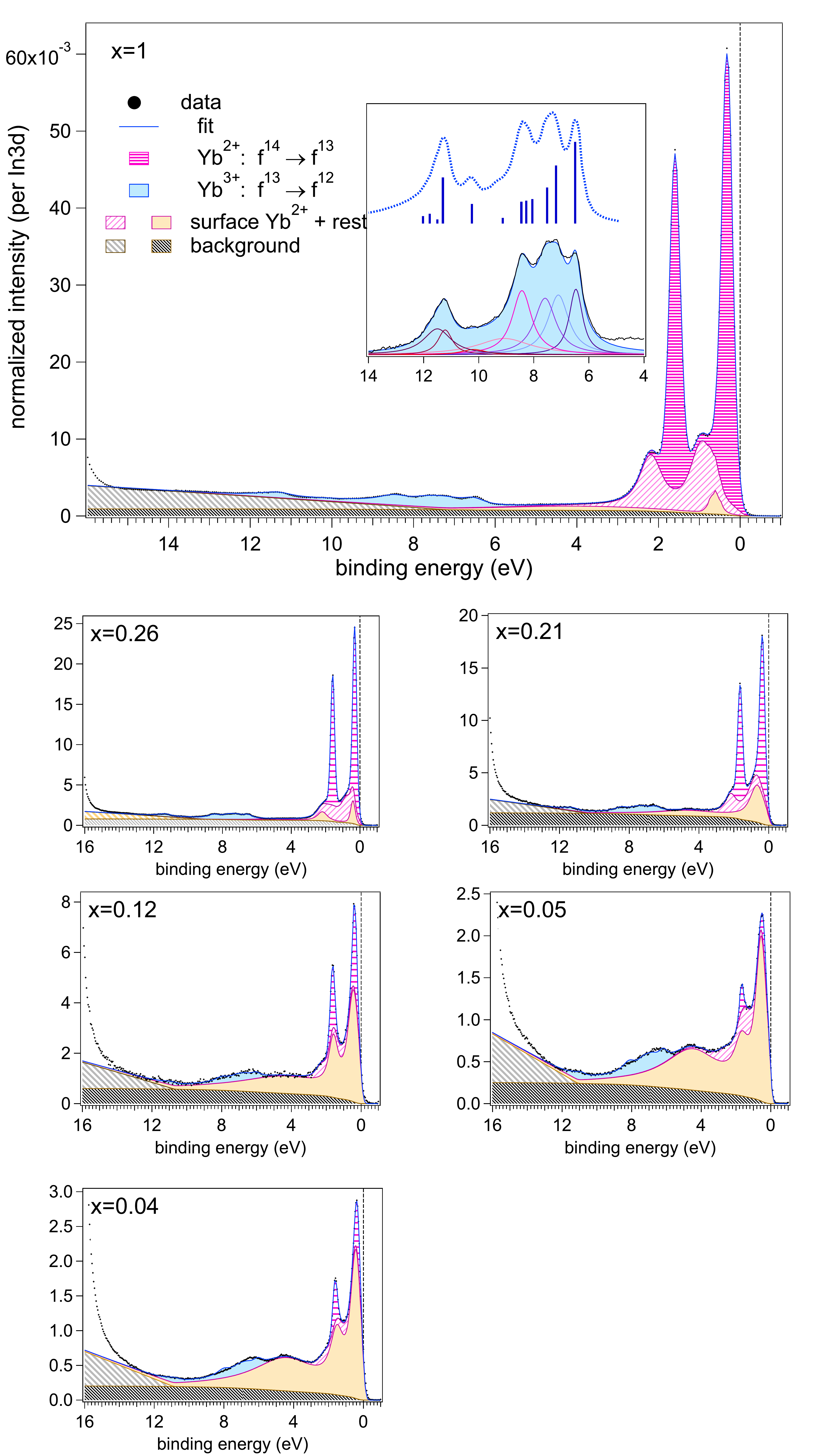}
  \caption{Example of the line-fits performed in order to determine the Yb-valence.
  The spectra shown are the same as in Fig. \ref{fig:XPS} (a), at \hv~$=$ 865\,eV. For the fit,
  we divided the spectrum into components from Yb$^{2+}$, surface
  Yb$^{2+}$ and Yb$^{3+}$. Furthermore, we use a Shirley background and a linear
  background and some subsequently added Lorentzians, just enough to optimally fit the spectrum (denoted as 'rest').
  The inset of $x$~$=$~1 shows the 7 Lorentzians we used to model the the twelve Yb$^{3+}$ 5$f^{13}$ \go\ 5$f^{12}$ lines.  }
  \label{fig:XPSFit}
  \end{center}
\end{figure}

In order to quantitatively extract the Yb valence, we applied a
fitting procedure presented in Fig. \ref{fig:XPSFit}. The background
used was a Shirley-like background and, additionally, a linear
background which simulates the contribution coming from the In 4$d$ and Ce 5$p$
doublets at higher binding energies. We modeled the F$_{5/2}$ -
F$_{7/2}$ doublet of the bulk Yb$^{2+}$  as two Lorentzians with the
fixed intensity ration 6:8. The Yb$^{2+}$ surface component was
modeled similarly. We reduced the twelve Yb$^{3+}$ 5$f^{13}$ \go\
5$f^{12}$ lines to 7 Lorentzians. The weights and positions of these
Lorentzians relative to each other were determined at $x$~$=$~1 (see inset
of $x$~$=$~1 in Fig. \ref{fig:XPSFit}). For fitting the other Yb
concentrations, we allow only the relative intensities of the two
valence components to change. The fitting routine first finds the
bulk Yb$^{2+}$-component and the Yb$^{3+}$ component together with
the background, adds then the surface Yb$^{2+}$, and after that
subsequently adds just enough extra peaks to optimally fit the
spectrum. These extra peaks are mainly to simulate the contributions
of Co-$d$ and Ce-$f$. For these extra peaks, we took two or three
Lorentzians. All spectral features were convolved with a Gaussian
having the FWHM of the resolution. Having so many components for a
line-fit, we may not always correctly distinguish between the
surface Yb$^{2+}$-component and these extra peaks. However, as can
be seen in Fig. \ref{fig:XPSFit}, the peaks labeled as 'rest'
represent very much what would be expected from the Co and Ce
weights. For these Ce weights, the reader can compare with Fig.
\ref{fig:XPS} (c).

\begin{figure}[h]
  \begin{center}
  \includegraphics[width=0.4\textwidth]{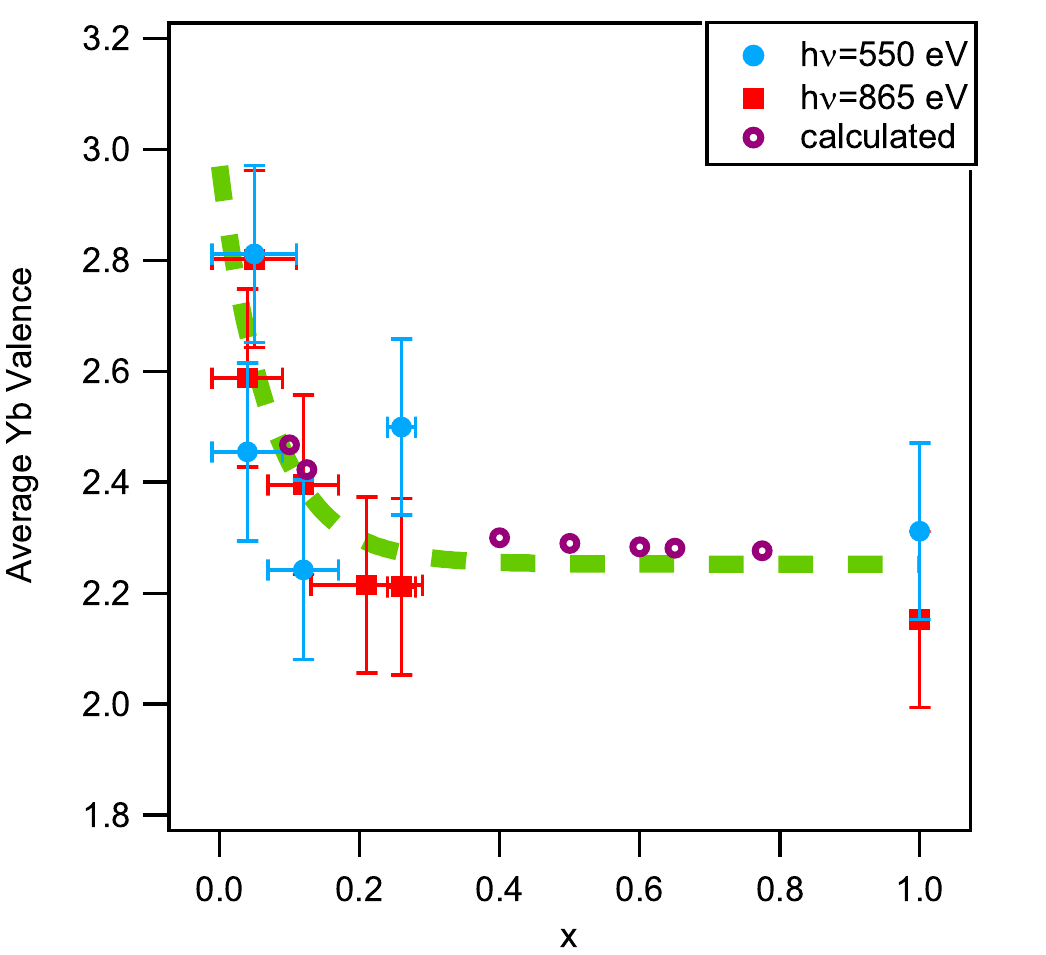}
  \caption{$x$-dependence of Yb valence from the XPS spectra at photon energies of h$\nu$~$=$~550\,eV and h$\nu$~$=$~865\,eV. The line is a guide to the eye. Open circles are the calculated Yb valences from the susceptibility analysis of the next section using Eq.~\ref{eq:v_yb}.}
  \label{fig:YbValence}
  \end{center}
\end{figure}

The result of the fitting procedure is summarized in Fig.
\ref{fig:YbValence}. There, we plot the fitting result for both
photon energies. The fact that the results are essentially the same for
both photon energies assures us that the fit is finding the amount of
the surface Yb$^{2+}$-component correctly and that the valence obtained
here reflects that of the bulk. For the general trend of the
valence vs. Yb concentration we see that the Yb valence is near 3+
at very low $x$ and goes down to about +2{.}3, as already concluded
qualitatively from Fig. \ref{fig:XPS}. Compared to the
results of Ref. \onlinecite{Booth2011}, we obtain the same valence
of about +2{.}3 for the end-member $x$~$=$~1. As discussed in Ref. \onlinecite{Booth2011}
the intermediate valence of the end-member together with the
nonmagnetic behavior seen in its magnetic susceptibility \cite{Shu2011}
indicates that \YbCo\ has a very high single ion Kondo temperature \TK.  However, by measuring for lower $x$-values than in
Ref. \onlinecite{Booth2011}, we obtain the new result that the Yb valence is strongly
increasing to trivalent for small values of $x$ going to zero. We will see in the next section that an
analysis of the valence from bulk properties is consistent
with the rather abrupt increase of valence at low $x$ observed spectroscopically. As noted already in Section ~\ref{subsec:ecounting} this change of Yb valence is also in good agreement with the change of electronic structure observed in dHvA experiments \cite{Polyakov2012}.

\section{Relation of Ce and Yb valences to bulk properties}\label{sec:bulk}

\begin{figure}[h]
  \begin{center}
  \includegraphics[width=0.45\textwidth]{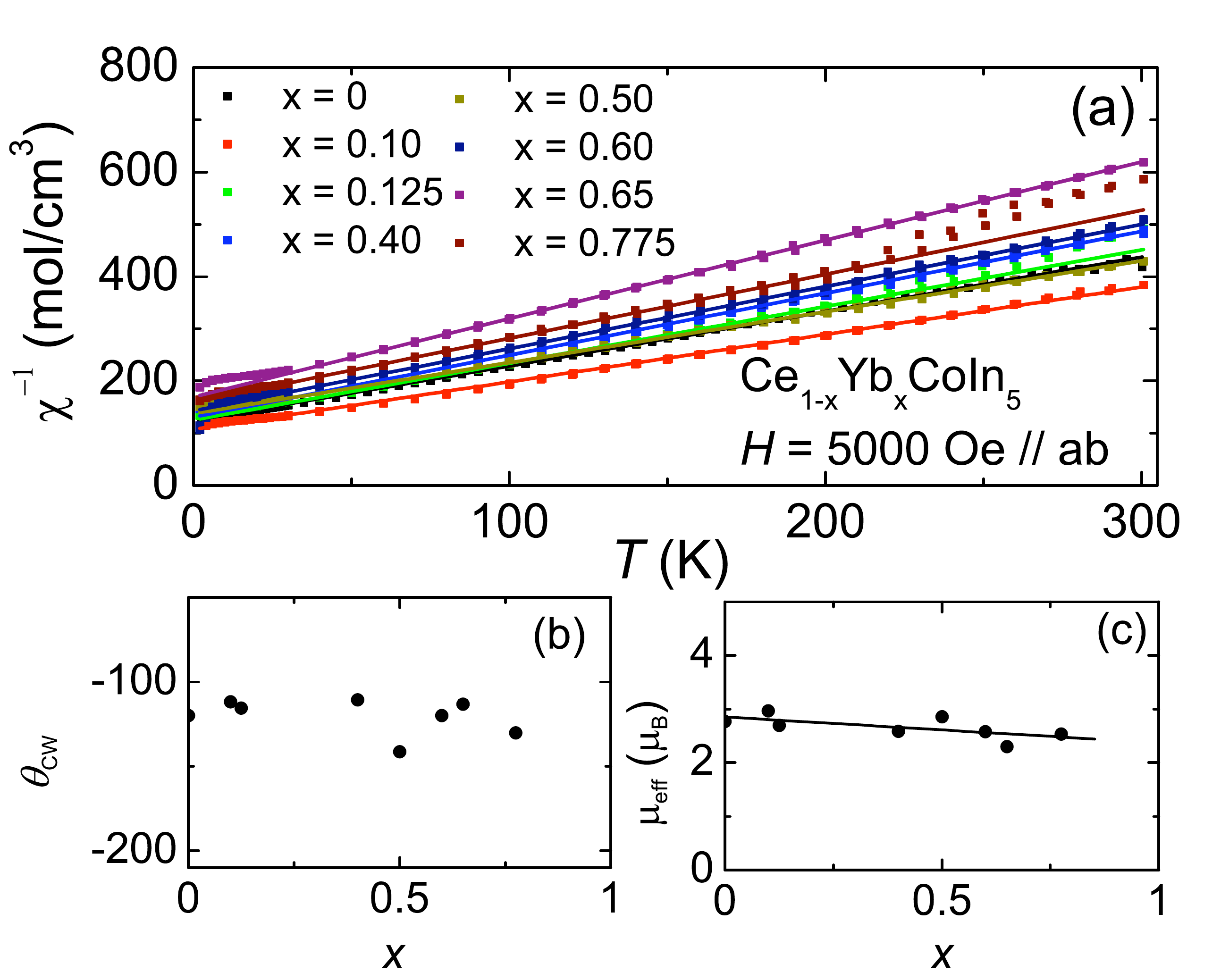}
  \caption{(a) Inverse magnetic susceptibility $\chi_{ab}^{-1}(T) = H/M_{ab}(T)$ vs.
  temperature $T$ along the crystallographic $ab$ plane for Ce$_{1-x}$Yb$_{x}$CoIn$_{5}$
  with Yb concentrations $0.0\leqslant x \leqslant 0.775$. Curie-Weiss temperature
  $\theta_{\rm CW}$ and effective magnetic moment $\mu_{\rm eff}$ as determined from
  fits to a Curie-Weiss law $\chi_{ab}=N_{A}\mu_{\rm eff}^{2}/3 k_{B}(T-\theta_{\rm CW})$(please see text)
  as function of Yb concentration $x$ are shown in (b) and (c), respectively.
  Solid line in (c): linear fit of $\mu_{\rm eff}(x)$.}
  \label{fig:chi}
  \end{center}
\end{figure}

Figure~\ref{fig:chi}(a) shows the inverse magnetic susceptibility
$\chi_{ab}^{-1}(T) = H/M_{ab}(T)$ of Ce$_{1-x}$Yb$_{x}$CoIn$_{5}$ in
the normal state vs. Yb concentration $x$. The data were collected
by warming the sample gradually after zero-field cooling (ZFC), and
subsequently field cooling (FC). The difference between the ZFC and
FC data is negligible. Above $T$ \~ 30 K and for all $x$, the magnetic
susceptibility $\chi_{ab}$ of Ce$_{1-x}$Yb$_{x}$CoIn$_{5}$
can be described well by a Curie-Weiss law
\begin{equation}
 \chi_{ab}=N_{A}\mu_{\rm eff}^{2}/3 k_{B}(T-\theta_{\rm CW}),\label{eq:CWlaw}
\end{equation}
where $N_{A}$ is Avogadro's number and $k_{B}$ is Boltzmann's
constant. The effective magnetic moment $\mu_{\rm eff}$ and the
Curie-Weiss temperature $\theta_{\rm CW}$ as determined from fits of
the data (solid lines  in Fig.~\ref{fig:chi}(a)) to
Eq.~\ref{eq:CWlaw} are shown in Fig.~\ref{fig:chi}(b) and (c).
The fits yield $\theta_{\rm CW}$~$\approx$~$-120$~K, independent of Yb concentration to first
approximation (Fig.~\ref{fig:chi}(b)). Curie-Weiss behavior is also
observed in temperature dependence of magnetic susceptibility
$\chi_{c}$ at high temperatures ($T\geqslant 50$~K), and the fits
give similar values of $\mu_{\rm eff}$ and $\theta_{\rm CW}$.

The magnetic susceptibility of metals containing lanthanide ions that exhibit the Kondo effect or valence fluctuations can be described by a Curie-Weiss law (Eq. \ref{eq:CWlaw}) in the high temperature limit.  For the present situation, the effective magnetic moments of the Ce and Yb ions are expected to be close to their Hund's rules values corresponding to their f-electron configurations and the Curie-Weiss temperature represents a characteristic temperature associated with the Kondo effect for the trivalent Ce ions or valence fluctuations for the intermediate valent Yb ions \cite{Maple1971,Maple1975}.  The Kondo and valence fluctuation temperatures are characteristic temperatures where the material gradually crosses over from a paramagnetic localized moment regime at high temperatures to a nonmagnetic Pauli-like regime at low temperatures.  In our analysis, the magnetic susceptibility is taken to be a superposition of a Kondo contribution from the Ce ions and a valence fluctuation contribution from the Yb ions.  The fact that Curie-Weiss temperature theta CW is nearly independent of Yb concentration indicates that the energy scale associated with the combined Kondo and valence fluctuation contributions does not vary with Yb concentration, which is consistent with the stability of the correlated electron state in \CeYbCo\ over a large Yb concentration range.

The magnetic susceptibility $\chi(x)_{ab}$ of
Ce$_{1-x}$Yb$_{x}$CoIn$_{5}$ is composed of two contributions
arising from Ce and Yb ions, respectively,
\begin{equation}
 \label{eq:chi}
\chi(x)_{ab}=\chi_{\rm Ce}(1-x)+\chi_{\rm Yb}x.
\end{equation}
Using the result that $\theta_{\rm CW}$ is independent of  $x$ (cf.
Fig.~\ref{fig:chi}(b)), we can write
\begin{equation}
 \label{eq:mu}
\mu_{\rm eff}^{2}(x)=\mu_{\rm Ce}^{2}(x) (1-x)+\mu_{\rm Yb}^{2}(x) x.
\end{equation}

Following the conclusion of the previous section and of
Ref. \onlinecite{Booth2011} we take the $f$-electron
orbital occupancy for Ce to remain close to 1 ($n_{f}^{\rm Ce}$\~ 1),
\textit{i.e}., the Ce ions remain $3+$ for all $x$. Thus, by assuming
that $\mu_{\rm Ce}(x)=\mu_{\rm Ce3+}=2.54\, \mu_{\rm B}$ for all $x$
and approximating $\mu_{\rm eff}(x)$ by the linear fit illustrated
in Fig.~\ref{fig:chi}(c), we can estimate $\mu_{\rm Yb}(x)$ by using
Eq.~(\ref{eq:mu}). The $f$-hole occupancy for Yb is given by
$n_{f}^{\rm Yb}(x)=\frac{\mu_{Yb}^{2}(x)}{\mu_{\rm Yb3+}^{2}}$,
where $\mu_{\rm Yb3+}=4.54 \mu_{\rm B}$. Accordingly, the effective
valence is then obtained as
\begin{equation}
 \label{eq:v_yb}
 v_{\rm Yb}(x)=2+n_{f}^{\rm Yb}(x).
\end{equation}
As shown by the open circles in Fig.~\ref{fig:YbValence}, the result
is in surprisingly good agreement with the results of the XPS measurements,
considering the simplicity of analysis of the magnetic susceptibility.

The analysis of the magnetic susceptibility measurements on the \CeYbCo\ crystals used to estimate the Yb valence relies on the value of the Yb concentration $x$.  Most of the magnetic susceptibility measurements were made on samples for which the composition measured by XEDS was close to the nominal composition (within about 5\%).  Coupled with the uncertainty in the magnetic susceptibility measurements, we roughly estimate that uncertainties in Yb valence are of the order of 10\%.  Samples whose compositions were not measured by XEDS had values of magnetic susceptibility and \Tc\ that changed systematically with nominal composition, indicating that the actual compositions are close to the nominal values. This suggests that the XEDS measurements may not be a reliable method of estimating the bulk Yb concentration in this system for reasons that are not presently understood.  In addition, we again stress that the assumptions on which the analysis of the magnetic susceptibility measurements is based, and that were used to estimate the Yb valence, are, while reasonable, not rigorously justified.

There is a well-known correlation between lattice parameters and valence in f-electron materials \cite{Gschneidner61}.  For the system Ce$_{1-x}$Yb$_{x}$CoIn$_{5}$, which has a tetragonal crystal structure with basal plane and interplane lattice parameters $a$ and $c$, respectively, we consider the relationship between the unit cell volume $V = a^2c$ and the valences of the Ce and Yb ions.  The lattice parameters $a$ ($c$) for CeCoIn$_5$ and YbCoIn$_5$ determined from XRD measurements are 4.6012\,\AA~(7.5537\AA) and 4.5590\,\AA~(7.433\,\AA),~\cite{Shu2011}, respectively, and the valence of Ce and Yb as obtained by our photoemission measurements and calculations are \~ 3+ and \~ 2.3+, respectively.  This yields unit cell volumes of $V$ = 159{.}9 \AA$^3$~ for CeCoIn$_5$ and 154{.}5 \AA$^3$~ for YbCoIn$_5$. Assuming that Vegard's law applies to the unit cell volume $V$ as a function of $x$ and that the valence of Ce remains near 3+ for all values of $x$, $V(x)$ can be expressed as
\begin{widetext}
\begin{eqnarray}
V(x) &=& V_{Ce^{3+}}(1-x) + V_{Yb}\, x \nonumber\\
&=& V_{Ce^{3+}}(1-x) + \left[ V_{Yb^{3+}}  + \left[V_{Yb^{2+}} - V_{Yb^{3+}}\right] \left(1 - n_f^{Yb}(x)\right) \right]\, x \nonumber\\
&=& 159.9\,\textrm{\AA}^3(1-x) + \left[ 155.6\,\textrm{\AA}^3 - 3.6\,\textrm{\AA}^3 n_f^{Yb}(x) \right]\, x   \label{eq:v}.						
\end{eqnarray}
\end{widetext}

The values of the unit cell volumes used in this expression are $V_{Ce^{3+}}$ = 159.9 \AA$^3$~, $V_{Yb^{2+}}$ = 155.6 \AA$^3$~, and $V_{Yb^{3+}}$  = 152.0 \AA$^3$~, while $n_f^{Yb}(x)$ is the number of holes in the Yb 4f-electron shell ($n_f^{Yb}(x)$ = 1 for Yb$^{3+}$ and 0 for Yb$^{2+}$).  The value of $V_{Yb^{3+}}$ was estimated by interpolating the values of $V_{Ln^{3+}}$ from the neighboring LnCoIn$_5$ compounds with trivalent Ln ions ~\cite{Zaremba2003} to YbCoIn$_5$.  The value of $V_{Yb^{2+}}$ was estimated using the value of $V_{Yb^{3+}}$ and $V_{Yb^{2.3+}}$, inferred from the XAS data reported herein for the compound YbCoIn$_5$, using the term for $V_{Yb}$ in brackets in Eq. ~\ref{eq:v}, which yields
\begin{equation}
 \label{eq:vYb2}
V_{Yb^{2+}} = \left[ V_{Yb}(x) - V_{Yb^{3+}}\,n_f^{Yb}(x)\right]/\left[1 - n_f^{Yb}(x)\right].					
\end{equation}
Using $n_f^{Yb}$(1) = 0.3 for $x$ = 1, we obtain $V_{Yb^{2+}}$ = 155.6 \AA$^3$~ from Eq.~\ref{eq:vYb2}.

\begin{figure}[h]
 \begin{center}
 \includegraphics[width=0.45\textwidth]{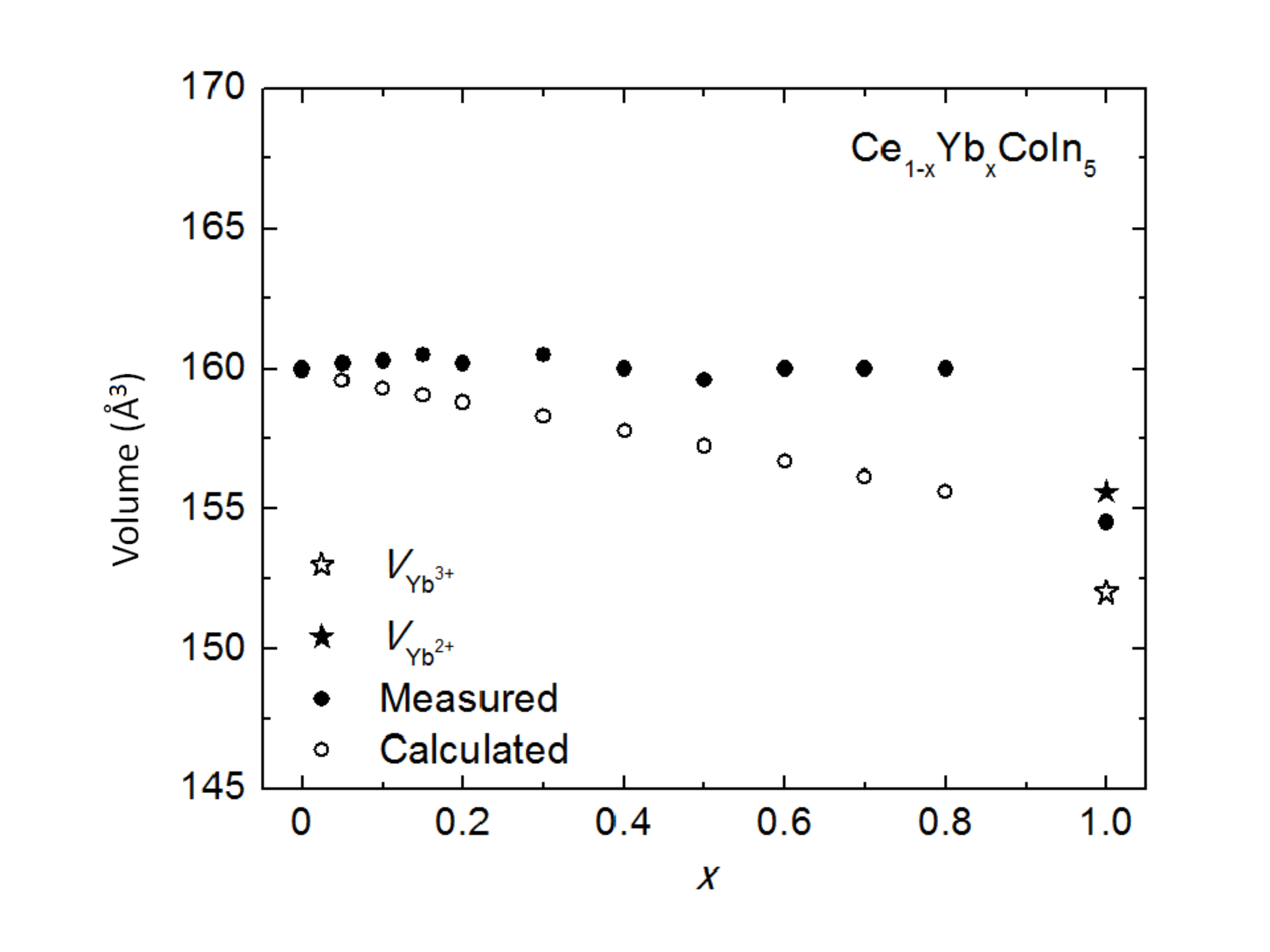}
 \caption{Unit cell volume $V$ as a function of Yb concentration $x$ in Ce$_{1-x}$Yb$_{x}$CoIn$_{5}$. Filled circles: measured values of $V$. Open circles: calculated $V$ using Eq.~\ref{eq:v}. Open circles: calculated $V$ using Eq.~\ref{eq:v}.}
 \label{fig:volume}
 \end{center}
\end{figure}

In Fig.~\ref{fig:volume}, we compare $V(x)$ determined from Eq.~\ref{eq:v} with the values obtained from the XRD measurements on Ce$_{1-x}$Yb$_{x}$CoIn$_{5}$.  It can be seen that the calculated values of $V(x)$ are nearly linear and generally conform to the behavior expected from Vegard's law for Ce$^{3+}$ and Yb$^{2.3+}$.  The calculated values of $V(x)$ are smaller than the measured values, and the discrepancy is larger for larger values of $x$.  A number of factors could contribute to this discrepancy; e.g., (1) Weakening of the metallic bond due to the decrease in the conduction electron density with Yb concentration that accompanies the substitution of Yb ions (valence \~ 2.3+) for Ce ions (valence \~ 3+), resulting in an increase of the unit cell volume, (2) a nonlinear contribution considered by Varma and Heine ~\cite{Varma1975} in calculating the unit cell volume for Ln compounds with intermediate valence, and (3) a reduction of the actual Yb concentration compared to the nominal concentration.

\section{Electronic Structure from ARPES}\label{sec:arpes}

In this section we present and discuss the rather columnar $\alpha$ and $\beta$ FS sheets for $x$~$=$~0, 0.2 and 1 as inferred from variable photon energy ARPES FS maps.  Thus far ARPES data of sufficiently high quality have not been obtained for other values of $x$.  In that connection we note that it was also challenging to obtain dHvA spectra for intermediate values of $x$, attributed in Ref. \onlinecite{Polyakov2012} to increased scattering rates in the alloys as inferred from Dingle temperatures that increased considerably from $x$~$=$~0 to $x$~$=$~0.55.  Such disorder would also degrade the sharpness of ARPES FS maps.

\begin{figure}[h]
  \begin{center}
  \includegraphics[width=0.45\textwidth]{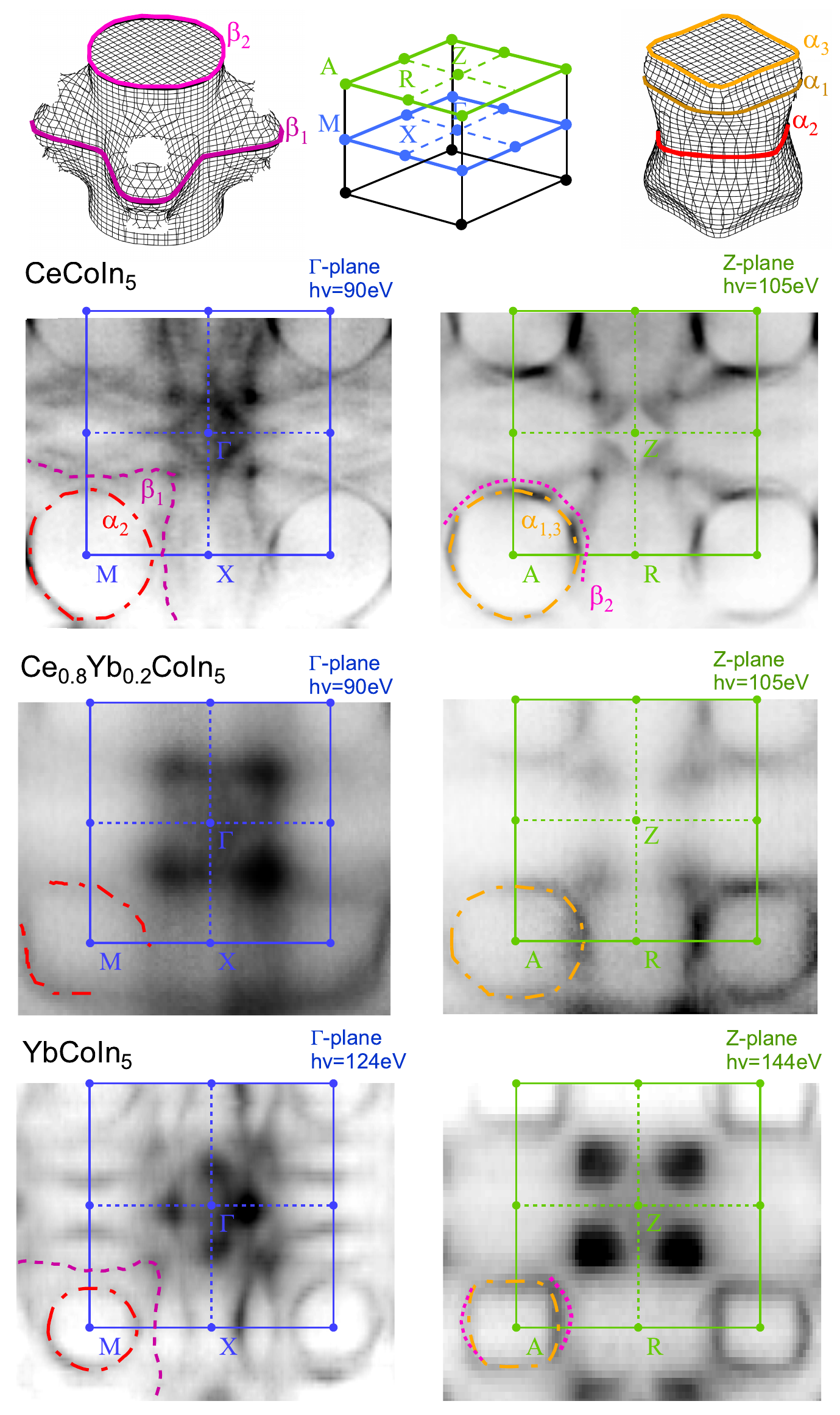}
  \caption{Second, third and and fourth rows show Fermi-surface maps for the \G and Z planes for $x$~$=$~0, 0.2 and 1 at 26\,K, 26\,K and 20\,K respectively.  Top row figures at right and left show respectively mesh models\cite{Shishido2003} of the columnar $\alpha$ and $\beta$-sheets as calculated in LDA, and middle figure shows the Brillouin zone and measurement planes.  The change of size of the $\alpha$ and $\beta$-sheets is easily seen.}
  \label{fig:maps}
  \end{center}
\end{figure}

The three lowest rows of Fig. \ref{fig:maps} show maps for $x$~$=$~0, 0.2 and 1 measured at 26 K, 26 K and 20 K,
respectively. Higher intensity correlates with darker color in these maps. We
show two interesting cuts though the high symmetry points of the
three dimensional Brillouin-zone (BZ). One cut contains the
$\Gamma$-point (left) and the other the Z-point (right). The
orientation of these two planes within the BZ is sketched in the
center of the upper panel of the figure.

There are many details visible in these six FS maps and furthermore
they represent only a small fraction of the data measured throughout
the whole BZ and in multiple zones. Here we will focus only on the $\alpha$- and
$\beta$-sheets which are readily identified in the data.  The more complex
FS pieces will be analyzed and discussed in a separate publication \cite{future}.  For \CeCo\, these complex pieces display topological differences that depend on whether or not\cite{Oppeneer07} the FS contains the Ce $f$-electron, \textit{i.e.}, whether the FS is "large" or "small."  For \CeCo\, ARPES finds \cite{JD,Koitzsch2009} that the Ce 4f electrons behave predominantly
localized for the present measurement temperature even though low $T$
de Haas van Alphen experiments \cite{Settai2001,Hall2001,Shishido2002}
unambiguously detect the large FS.  For the $\alpha$ and $\beta$-sheets
discussed here LDA calculations \cite{Oppeneer07} performed for isoelectronic \CeRh\ with the Ce $f$-electron confined to the core show only small changes in size and no dramatic topological changes due to localizing the $f$-electron and excluding it from the FS.  Nonetheless we will now see that these sheets display large size changes as $x$ varies from 0 to 1.  For $x$~$=$~0 and 1 these size changes are in good agreement with findings from dHvA experiments \cite{Polyakov2012} and accompanying LDA calculations for \YbCo.  However for the intermediate value of x there is an important difference from the dHvA results, as discussed below.

The two mesh models at the top right and left in the figure
show for $x$~$=$~0 the three dimensional shapes of the $\alpha$-sheet and
the $\beta$-sheet, respectively. The mesh models are based on an
itinerant LDA calculation and are taken from Ref. \onlinecite{Shishido2003}.
The $\alpha$-sheet is a nearly two-dimensional infinite cylinder
along the (001)-direction (k$_z$-direction). Measurements using the
de Haas van Alphen effect observe three distinct (001) frequencies
which LDA identifies as not only the $\Gamma$-plane diamond
($\alpha_2$) and Z-plane square ($\alpha_3$) orbits but also a slightly
larger circular orbit ($\alpha_1$) just above and below the Z-plane
\cite{Settai2001}. Comparing the mesh models with the ARPES measurements, we
identify the large round contours centered on the BZ corners as
coming from the quasi-two-dimensional $\alpha$-sheet. Consistent
with this identification, this contour evolves from a diamond-like
shape in the $\Gamma$-plane of \CeCo\ to a slightly squarish shape
in the Z-plane. Due to the experimental broadening of ARPES in the
k$_z$-direction, the observed Z-plane $\alpha$-sheet contour has a
distinctly broader width with a squarish inside edge and a rounder
outside edge, which is interpretable as the contour of $\alpha_3$ being blurred
with that of $\alpha_1$. For \YbCo, we can similarly identify the same shapes
as belonging to the $\alpha$-sheet.  The contours observed in the
Z-plane have more of a squarish shape than those of \CeCo, perhaps consistent with LDA calculations \cite{Polyakov2012}
for \YbCo\ that show a somewhat different $\alpha$ column shape.  The LDA calculations also show
corrugations for intermediate k-values that can give rise to two other orbits likely seen in dHvA
but not resolved in ARPES.  We next identify the $\beta$-sheet as the flower-like lobes in the
$\Gamma$-planes of \CeCo\ and \YbCo. Although the circular shape of the
$\beta$-sheet is harder to observe in the Z-plane, we note
small crescent-like pieces which are clearly visible aligned along
the AR-direction. The low intensity of the $\beta$-sheets along the
AZ-direction is most likely caused by a matrix element selection
rule.  For $x$~$=$~0.2 the data allow only an identification of
the $\alpha$ sheet cross-sections but not those of the $\beta$ sheet.

\begin{table*}[htb]
\begin{center}
\begin{tabular}{|c||c|c|c|c||c|c|c|}
  \hline

  & \multicolumn{4}{c||}{ARPES}& \multicolumn{3}{c|}{dHvA}\\
    Composition  & \multicolumn{4}{c||}{}& Ref. \onlinecite{Polyakov2012}& Ref. \onlinecite{Hall2001}& Ref. \onlinecite{Settai2001}\\\cline{2-8}
  & Label  & FS-Area & FS Area & $\Delta \nu$ & $\Delta \nu$ &$\Delta \nu$ & $\Delta \nu$\\
  &  & $(\pi/a)^2$ & rel. to x=0 & kT   & kT& kT& kT\\\hline\hline

   x=0& $\alpha_{1,3}$& 0.85 $\pm$ 0.06& 1 & 4.1 $\pm$ 0.3 & 5.46 ($\alpha_1$)/4.37 ($\alpha_3$)& 5.401 ($F_5$)/4.566 ($F_3$)& 5.56 ($\alpha_1$)/4.24 ($\alpha_3$) \\\cline{2-8}
   & $\alpha_{2}$ & 0.81 $\pm$ 0.06 & 1 & 3.9 $\pm$ 0.3& 4.87 ($\alpha_2$) & 5.161 ($F_4$)& 4.53 ($\alpha_2$)\\\cline{2-8}
    & $\beta_{1}$ & 1.8 $\pm$ 0.15& 1 & 8.7 $\pm$ 0.7 & 11.6 ($\beta_1$)$^{\dagger}$  & & 12.0 ($\beta_1$)\\\cline{2-8}
   &  $\beta_{2}$& 1.12 $\pm$ 0.15& 1 & 5.4 $\pm$ 0.7 & 7.4 ($\beta_2$)$^{\dagger}$ & 7.535 ($F_6$) & 7.5 ($\beta_2$) \\\hline\hline

   x=0.2& $\alpha_{1}$& 0.77 $\pm$ 0.15& 0.91 &  3.7 $\pm$ 0.7& & &\\\cline{2-5}
   & $\alpha_{2}$ & 0.74 $\pm$ 0.2& 0.91 &  3.6 $\pm$ 1& & &\\\hline\hline

    x=1& $\alpha_{1}$& 0.46 $\pm$ 0.08& 0.54 &  2.2 $\pm$ 0.4& 2.2 ($F_5$)$^{\dagger}$& &\\\cline{2-6}
  & $\alpha_{2}$ & 0.38 $\pm$ 0.06& 0.46 & 1.8 $\pm$ 0.3 &  1.6 ($F_3$)$^{\dagger}$& &\\\cline{2-6}
    & $\beta_{1}$ & 1.1 $\pm$ 0.08 & 0.61 & 5.3 $\pm$ 0.4 & 6.84 ($\beta_2$/$F_9$)& &\\\cline{2-6}
   &  $\beta_{2}$& 0.5 $\pm$ 0.1& 0.44 & 2.4 $\pm$ 0.5 & 3.66 ($\beta_1$/$F_8$)& &\\\hline

\end{tabular}
\end{center}
\caption{Left portion:  Fermi-surface cross-sectional areas and implied dHvA frequencies from ARPES Fermi-surface maps calculated in units of kilo Tesla (kT);  Right portion:  Measured dHvA frequencies with original paper labeling shown in parentheses.  For $x$~$=$~1 Ref.  \onlinecite{Polyakov2012} reverses the labeling of the $\beta$-sheet orbits from that used for $x$~$=$~0 such that $\beta_{1}$ corresponds to the Z-plane and $\beta_{2}$ corresponds to the $\Gamma$-plane.
$^{\dagger}$Value extracted from a publication graph because not explicitly tabulated.}\label{table-B}
\end{table*}

Comparing $x$~$=$~0 to 0.2 to 1, the noteworthy change clearly visible by eye in the FS maps is
 the considerable reduction of size for the $\alpha$-sheet and the
$\beta$-sheet. In contrast to the conclusion of Ref. \onlinecite{Booth2011}, our
results show that the electronic structure along the \G -M line is much different for $x$~$=$~0 and $x$~$=$~1, even though the general shapes of the $\alpha$ and $\beta$-sheets are much the same.  We have made a quantitative analysis to determine the cross-section areas in units of ($\pi$/a)$^2$, along with the implied dHvA frequencies.  The results are listed in Table II along with measured dHvA frequencies for $x$~$=$~0 and 1.  As discussed below a direct comparison to dHvA frequencies for $x$~$=$~0.2 is not possible.  For $x$~$=$~0 there is good general agreement among the dHvA frequencies for the three studies cited.  The ARPES frequencies have a similar pattern of variation in magnitude among the various orbits but are systematically smaller than those from dHvA, a difference for which one can consider two contributing effects.  The first is the difference in temperature of the two measurements, that low-$T$ dHvA sees the "large" FS and that higher-$T$ ARPES likely sees the "small" FS or at least a smaller FS.  However dHvA \cite{Settai2001,Hall2001,Shishido2002} and LDA \cite{Oppeneer07} studies find that the fractional differences occurring in the $\alpha$ and $\beta$ sheets for the change from localized to itinerant Ce f electrons in the Ce 115 compounds are relatively small, so this effect may not be very important.  Second, even at the ARPES measurement temperature the bands are still heavy enough very near \EF\ over the energy window of the FS maps that the \kF\ value for an electron pocket is likely to be underestimated.   For $x$~$=$~1 neither of these differences are important because the measurement temperatures for dHvA and ARPES are both much less than the characteristic temperature and the bands are light.  Indeed here we find much better quantitative agreeement between the two techniques.  It should however be noted that for the $\alpha$-sheet the good agreement is aided by assigning the measured dHvA frequencies somewhat differently than in Ref.  \onlinecite{Polyakov2012}.  In that paper the four frequencies labeled $F_3$ through $F_6$ (1.8 kT, 2.04 kT, 2.19 kT and 2.96 kT, respectively) are all associated with the $\alpha$-sheet but $F_4$ is assigned to $\alpha_{1}$ and $F_6$ is assigned to $\alpha_{2}$, leaving $F_3$ and $F_5$ to be assigned to the corregations predicted in LDA between the $\Gamma$-plane and the Z-plane.   Considering the shape of the LDA $\alpha$-sheet we find it more natural and in better agreement with ARPES to assign $F_4$ as $\alpha_{1}$ and $F_6$ as $\alpha_{2}$, with the other two belonging to the corregations.  These reassigments are not inconsistent with the observed angular dependences of the orbits.

Table II also lists values for the ARPES FS areas for $x$~$=$~1 and $x$~$=$~0.2 relative to those for $x$~$=$~0.  For the $\Gamma$-plane, the ratio between the $\alpha$-sheet area in \YbCo\ compared to the area in \CeCo\ is 0.46, while for the Z-plane this ratio is 0.54.  From these data we would estimate that the average ratio of the volumes of the $\alpha$-sheets is rougly 50\%.  We note that Ref.  \onlinecite{Polyakov2012} also concluded a value of about 50\% from the dHvA data, but the combination of the larger dHvA frequencies for $x$~$=$~0 and our changed orbit assignments for $x$~$=$~1, as discussed in the preceding paragraph, causes the dHvA data of Table II to imply a ratio somewhat smaller, perhaps 40\%.  The ratio of the volumes of the $\beta$-sheets is estimated to be about 50\% from the ARPES and dHvA data.  By the electron counting discussion in Section ~\ref{subsec:ecounting}, one would expect for ARPES the total FS volume of \YbCo\ compared to that of \CeCo\ to include 2 electrons/Yb compared to 3 electrons/Ce which results in a ratio of 66\%, whereas for dHvA the FS volume would change from including 2 electrons/Yb to 4 electrons/Ce which results in a ratio of 50\%.  Considering the uncertainties discussed in the preceding paragraph, and that we can not expect any particular part of the FS to change in strict proportion to the change of the total, these findings are very consistent with the general expectations of simple electron counting.

The significant finding for the ARPES data for $x$~$=$~0.2 is that the $\alpha$-sheet areas are, even by eye, intermediate between those for $x$~$=$~0 and $x$~$=$~1.  Quantitatively the change in area from $x$~$=$~0 to $x$~$=$~0.2 is 21\% and 14\% of the change from $x$~$=$~0 to $x$~$=$~1 for $\alpha_{1}$ and $\alpha_{2}$ respectively, i.e. roughly in the same proportion as the doping.  Unfortunately it is not possible to compare this result with the dHvA data for intermediate values of $x$.  Although the frequency $F_7$  of the dHvA data is very similar to that implied by the ARPES data (i.e., 3.6 kT or 3.7 kT) this association cannot be made because $F_7$ has hardly any variation with x and in Ref.  \onlinecite{Polyakov2012} is thought to be associated somehow with the $\beta$-sheet.   Indeed the dHvA frequencies associated with the $\alpha$-sheets and $\beta$-sheets of the end members $x$~$=$~0 and $x$~$=$~1 show only slight modification for intermediate $x$-values, quite different from the ARPES finding here of a clear intermediate $\alpha$-sheet size for an intermediate $x$-value.  We have no hypothesis for this very great difference beyond the possibility of some sample difference for intermediate $x$.


\section{Summary and conclusions}\label{sec:conclusion}

There are three main new findings of this work for \CeYbCo.  First, for small values of $x$ increasing from 0, the Yb valence changes rapidly from being nearly trivalent to the value of 2.3 found previously for larger $x$ and for \YbCo.  Second, we have  directly observed a large reduction in the sizes of the columnar $\alpha$ and $\beta$ FS features for $x$~$=$~1 relative to those for $x$~$=$~0.   As already noted, both of these findings are in very good agreement with dHvA results \cite{Polyakov2012}.  Taken together, these results imply that around $x$~$=$~0.2 a change of Yb valence drives a switch of the near \EF\ electronic structure from one characteristic of $x$~$=$~0 with some very heavy mass FS pieces to one characteristic of $x$~$=$~1 with Yb valence around 2.3, a very large Yb \TK\ and small measured masses.  Third, for at least one intermediate $x$-value and one sheet ($\alpha$), the FS has been found to evolve between that of the two end memebers.    This third result contrasts quite sharply with the dHvA results, in which the observed frequencies and masses do not evolve significantly with $x$ and change quickly from one to the other, with a mix of both being seen for $x$~$=$~0.2.

What are the implications of these findings for understanding the transport properties?  Regarding the two hypotheses mentioned in Section ~\ref{subsect:GenTrans}, that of cooperative valence fluctuations or coexisting networks of \CeCo\ and \YbCo,  the ARPES results clearly favor the former, at least for the samples used in this study.   The addition of only a small amount of Yb drives an overall change in the FS and near \EF\ electronic structure, and there appear to be unified electronic structures presumably involving the f-states of both Ce and Yb.  On the other hand the dHvA results lead to a different conclusion.  The finding there of aspects of both electronic structures at $x$~$=$~0.2 suggests some coexistence.  Further it is very puzzling that the observed dHvA frequencies, e.g., for the $\alpha$ orbits,  do not change as $x$ increases beyond 0.2, given that the FS size changes appear to be driven by simple electron counting and that the total number of electrons to be contained in the FS per rare earth atom certainly changes steadily.  Also, as noted in Ref. \onlinecite{Polyakov2012}, the switch to a FS with only low mass measured features is inconsistent with the finding that the specific heat is roughly constant with $x$.  This issue gave rise to the conclusion \cite{Polyakov2012} that there must be heavy FS pieces for larger $x$ not yet observed in dHvA, possibly the heavier $\beta$ orbits that were not observed for $x$~$=$~0.2 and 0.55.   The change in transport properties across the crystallographically two-phase region between $x$~$=$~0.8 and $x$~$=$~1 also bears thought.  The absence of Kondo-like features in the resistivity for $x$~$=$~1 is readily understandable from the Yb valence of 2.3, the implied large \TK\ and the low dHvA masses.  But the presence of these features for $x$~$=$~0.8, very similar to those for $x$~$=$~0, plus the lack of any Yb valence change across the region, implies that the resistivity is due only to the Ce and that the change is simply the result of removing all the Ce from the lattice.  In this respect the Ce would seem to be acting independently of the Yb.

What is the role of the Ce and Yb for the SC?  That \Tc\ decreases only slowly and gradually with $x$ implies a gradual steady change of some essential ingredients for the SC and it is again tempting to think of Ce and Yb as acting independently.    One possibility is that Ce brings a local moment which is essential for the SC.  In this picture the smallness of the Ce \TK\ is a benefit, and Yb, with the largeness of its \TK\,  does no direct harm but does serve to dilute the Ce.  Another model involving Ce as the essential active ingredient perhaps being diluted by Yb is the composite pairing picture \cite{Flint}.  Moving away from local pictures, if low dimensionality is involved in the SC, as has been suggested for pairing involving spin fluctuations, then the particulars of the columnar pieces of FS might be important. The changing sizes of these pieces with $x$ in the ARPES data (but not the dHvA data) would be consistent with this idea.

In conclusion, while good progress has been made in determining the electronic structure of this interesting new alloy series, it remains to find a unified view that explains both what is now known about the electronic properties and what is known about the transport properties.  One step forward for the future would be to obtain a complete set of ARPES data for intermediate values of $x$, ideally having the same high quality as that reported here for $x$~$=$~0 and $x$~$=$~1.

\section{Acknowledgement}

Supported by the U.S. DOE at the ALS, Contract No.
DE-AC02-05CH11231, at UM, Contract No. DE-FG02-07ER46379 for current
work, and at UCSD, Contract No. DE FG02-04ER-46105; by the U.S. NSF
at UM, Grant No. DMR-03-02825 for initial work.  The experimental
support at ALS beamline 7.0 by E. Rotenberg is gratefully
acknowledged.  For the EDX measured at the EMAL at the UM, we thank
J. Mansfield for discussion and acknowledge the support of NSF grant
DMR-0320740. MJ acknowledges support by the Alexander von Humboldt
foundation. We thank Sooyoung Jang for assistance in the preparation of Fig. \ref{fig:volume}.

\end{document}